\documentclass[5p,authoryear]{elsarticle}

 \usepackage{graphicx}
 \usepackage{subfigure}
\usepackage{cleveref}

\usepackage{multirow}
\usepackage{array}

\newcolumntype{M}[1]{>{\centering\arraybackslash}m{#1}}
\newcolumntype{N}{@{}m{0pt}@{}}
\usepackage{amssymb}
\usepackage{amsmath}
\usepackage{amsfonts}
\journal{Neuroimage}

\begin{document}

\begin{frontmatter}

%% Title, authors and addresses

%% use the tnoteref command within \title for footnotes;
%% use the tnotetext command for the associated footnote;
%% use the fnref command within \author or \address for footnotes;
%% use the fntext command for the associated footnote;
%% use the corref command within \author for corresponding author footnotes;
%% use the cortext command for the associated footnote;
%% use the ead command for the email address,
%% and the form \ead[url] for the home page:
%%
%% \title{Title\tnoteref{label1}}
%% \tnotetext[label1]{}
%% \author{Name\corref{cor1}\fnref{label2}}
%% \ead{email address}
%% \ead[url]{home page}
%% \fntext[label2]{}
%% \cortext[cor1]{}
%% \address{Address\fnref{label3}}
%% \fntext[label3]{}

\title{Nonlinear functional mapping of the human brain}

%% use optional labels to link authors explicitly to addresses:
%% \author[label1,label2]{<author name>}
%% \address[label1]{<address>}
%% \address[label2]{<address>}

\author[uvmmath,csc,uvmcs,uvmpsych]{Allgaier, N.A.\corref{cor}}%\fnref{fn}}
\cortext[cor]{Corresponding author.}
\ead{nicholas.allgaier@uvm.edu} 
%\fntext[fn]{+1 508 292 3489}
%\author[10]{Artiges, E.} 
\author[1]{Banaschewski, T.}
\author[2]{Barker, G.J.} 
\author[3]{Bokde, A.L.W.} 
\author[csc,uvmcs]{Bongard, J.C.}
\author[4]{Bromberg, U} 
\author[4]{B\"uchel, C.}
\author[5]{Cattrell, A.} 
\author[6,7]{Conrod, P.} 
\author[uvmmath,csc]{Danforth, C.M.}
\author[5]{Desrivi\`eres, S.}
\author[uvmmath,csc]{Dodds, P.S.}
\author[8]{Flor, H.} 
\author[9]{Frouin, V.} 
\author[10]{Gallinat, J.} 
\author[12]{Gowland, P.} 
\author[13]{Heinz, A.} 
\author[14]{Ittermann, B.} 
%\author[18]{Lemaitre, H.} 
%\author[1,2]{Loth, E.} 
\author[uvmpsych]{Mackey, S.}
%\author[20]{Mann, K.} 
\author[15]{Martinot, J-L.} 
\author[24]{Murphy, K.}
%\author[10]{Miranda, R.} 
\author[8]{Nees, F.} 
%\author[10]{Paillere-Martinot, M.L.} 
\author[9]{Papadopoulos-Orfanos, D.}
%\author[19]{Paus, T.} 
%\author[19]{Pausova, Z.} 
%\author[15]{Poline, J.B.}
\author[1,20]{Poustka, L.} 
%\author[3,4]{Rietschel, M.} 
%\author[16]{Robbins, T.} 
\author[21]{Smolka, M.N.} 
%\author[7]{Str\"ohle, A.}
%\author[21]{Vetter, N.C.}
%\author[10]{Vulser, H.}
\author[13]{Walter, H.} 
\author[23]{Whelan, R.}
\author[5]{Schumann, G.} 
\author[uvmpsych]{Garavan, H.} 
\author{IMAGEN Consortium}
\ead[url]{www.imagen-europe.com}

\address[uvmmath]{Department of Mathematics and Statistics, University of Vermont, Burlington, Vermont, USA}
\address[csc]{Vermont Complex Systems Center, Vermont Advanced Computing Core, Burlington, Vermont, USA}
%\address[vacc]{Vermont Advanced Computing Core, Burlington, VT, USA}
\address[uvmcs]{Department of Computer Science, University of Vermont, Burlington, Vermont, USA}
\address[uvmpsych]{Departments of Psychiatry and Psychology, University of Vermont, 05405 Burlington, Vermont, USA}

\address[1]{Department of Child and Adolescent Psychiatry and Psychotherapy, Central Institute of Mental Health, Medical Faculty Mannheim, Heidelberg University, Square J5, 68159 Mannheim, Germany}
\address[2]{Centre for Neuroimaging Sciences, Institute of Psychiatry, Psychology \& Neuroscience, KingÕs College London, United Kingdom}
\address[3]{Discipline of Psychiatry, School of Medicine and Trinity College Institute of Neurosciences, Trinity College Dublin, Ireland} 
\address[4]{University Medical Centre Hamburg-Eppendorf, House W34, 3.OG, Martinistr. 52, 20246, Hamburg, Germany}
\address[5]{Medical Research Council - Social, Genetic and Developmental Psychiatry Centre, Institute of Psychiatry, Psychology \& Neuroscience, KingÕs College London, United Kingdom}
\address[6]{Department of Psychiatry, Universit\'e de Montreal, CHU Ste Justine Hospital, Canada} 
\address[7]{Department of Psychological Medicine and Psychiatry, Institute of Psychiatry, Psychology \& Neuroscience, King's College London}

\address[8]{Department of Cognitive and Clinical Neuroscience, Central Institute of Mental Health, Medical Faculty Mannheim, Heidelberg University, Square J5, Mannheim, Germany} 
\address[9]{Neurospin, Commissariat \`a l'Energie Atomique, CEA-Saclay Center, Paris, France} 
\address[10]{Department of Psychiatry and Psychotherapy, University Medical Center Hamburg-Eppendorf (UKE), Martinistrasse 52, 20246 Hamburg} 
%\address[11]{Departments of Psychiatry and Psychology, University of Vermont, 05405 Burlington, Vermont, USA} 
\address[12]{Sir Peter Mansfield Imaging Centre School of Physics \& Astronomy, University of Nottingham, University Park, United Kingdom} 
\address[13]{Department of Psychiatry and Psychotherapy, Campus Charit\'e Mitte, Charit\'e, Universit\"atsmedizin Berlin, Germany} 
\address[14]{Physikalisch-Technische Bundesanstalt (PTB), Braunschweig and Berlin, Germany} 
\address[24]{Cardiff University Brain Research Imaging Centre (CUBRIC), School of Psychology, Cardiff University, Cardiff, United Kingdom}
\address[15]{Institut National de la Sant\'e et de la Recherche M\'edicale, INSERM Unit 1000 ÒNeuroimaging \& PsychiatryÓ, University Paris Sud, University Paris Descartes - Sorbonne Paris Cit\'e; and Maison de Solenn, Paris, France} 
%\address[16]{Institut National de la SantŽ et de la Recherche MŽdicale, INSERM Unit 1000 ÒNeuroimaging & PsychiatryÓ, University Paris Sud, University Paris Descartes - Sorbonne Paris CitŽ; and AP-HP, Department of Adolescent Psychopathology and Medicine, Maison de Solenn, Cochin Hospital, Paris, France}
%\address[17]{Institut National de la SantŽ et de la Recherche MŽdicale, INSERM Unit 1000 ÒNeuroimaging & PsychiatryÓ, University Paris Sud, University Paris Descartes - Sorbonne Paris CitŽ; and Psychiatry Department 91G16, Orsay Hospital, France}
%\address[18]{Institut National de la SantŽ et de la Recherche MŽdicale, INSERM Unit 1000 ÒNeuroimaging \& PsychiatryÓ, FacultŽ de mŽdecine, UniversitŽ Paris-Sud, Le Kremlin-Bictre; and UniversitŽ Paris Descartes, Sorbonne Paris CitŽ, Paris, France}
%\address[19]{Rotman Research Institute, Baycrest \& Departments of Psychology \& Psychiatry, University of Toronto, Toronto, Ontario, Canada} 
\address[20]{Department of Child and Adolescent Psychiatry and Psychotherapy, Medical University of Vienna, Austria}
\address[21]{Department of Psychiatry and Neuroimaging Center, Technische Universit\"at Dresden, Dresden, Germany}
\address[23]{Department of Psychology, University College Dublin}

%\vspace{-3in}

\begin{abstract}

%!TEX root =  InteractionHierarchy.tex
The field of neuroimaging has truly become data rich, and novel analytical methods capable of gleaning meaningful information from large stores of imaging data are in high demand.  Those methods that might also be applicable on the level of individual subjects, and thus potentially useful clinically, are of special interest.  In the present study, we introduce just such a method, called \textsl{nonlinear functional mapping} (NFM), and demonstrate its application in the analysis of resting state fMRI (functional Magnetic Resonance Imaging) from a 242-subject subset of the IMAGEN project, a European study of adolescents that includes longitudinal phenotypic, behavioral, genetic, and neuroimaging data. NFM employs a computational technique inspired by biological evolution to discover and mathematically characterize interactions among ROI (regions of interest), without making linear or univariate assumptions. We show that statistics of the resulting interaction relationships comport with recent independent work, constituting a preliminary cross-validation. Furthermore, nonlinear terms are ubiquitous in the models generated by NFM, suggesting that some of the interactions characterized here are not discoverable by standard linear methods of analysis.  We discuss one such nonlinear interaction in the context of a direct comparison with a procedure involving pairwise correlation, designed to be an analogous linear version of functional mapping.  We find another such interaction that suggests a novel distinction in brain function between drinking and non-drinking adolescents: a tighter coupling of ROI associated with emotion, reward, and interoceptive processes such as thirst, among drinkers.  Finally, we outline many improvements and extensions of the methodology to reduce computational expense, complement other analytical tools like graph-theoretic analysis, and allow for voxel level NFM to eliminate the necessity of ROI selection.

%In addition, variation across subjects in ROI interaction allows for phenotypic differentiation. Such differentiation is demonstrated by the separation of low and high verbal, and performance IQ sub-populations.   Finally, we note that the statistical analysis employed here to interpret the collection of models generated by functional mapping is the coarsest possible.  Thus, the results we present likely just scratch the surface of the information contained in the output of the algorithm.  

\end{abstract}

\begin{keyword}

resting state fMRI \sep modeling \sep nonlinear \sep machine learning \sep genetic programming \sep symbolic regression
%% keywords here, in the form: keyword \sep keyword

%% MSC codes here, in the form: \MSC code \sep code
%% or \MSC[2008] code \sep code (2000 is the default)

\end{keyword}

\end{frontmatter}

% \linenumbers

%% main text
% !TEX root = InteractionHierarchy.tex
\section{Introduction}

Many advances in our understanding of brain function have been achieved through analysis of fMRI data.  Though the BOLD (blood oxygen level dependent) signal obtained from fMRI is a proxy, physiological confounds such as breathing and heart rate are separable from neuronal-induced signal, as demonstrated in \cite{birn2009}. Inter-subject differences in vascular reactivity can be modeled as shown in \cite{murphy2011}, and BOLD has been directly shown to provide a reliable measure of neuronal activity in specific circumstances, as in \cite{mukamel2005}. The many years of successful research before and since support that assessment.  Accomplishments include localization of regions responsible for particular tasks, such as episodic memory in \cite{nolde1998} and human face recognition in \cite{kanwisher1999}, assessment of the risk of postoperative motor defect in patients with tumors in \cite{mueller1996}, analysis of the effects of acupuncture in \cite{hui2000}, and recently, identification of neural markers for both current \textsl{and future} alcohol use among adolescents in \cite{whelan2012} and \cite{whelan2014}.  

These examples, and indeed the majority of fMRI studies, make use of the GLM (general linear model) to determine neural correlates for various tasks and stimulus responses. 
%Additionally, recent work reported in \cite{ide2013} investigated a hypothesized computational model of inhibitory control, and used the GLM to determine neural correlates of that \textsl{cognitive} function. 
Though typical analyses have been performed at the group level with a univariate approach, other recent work reported in \cite{rio2013} has extended the capabilities of the GLM to analyze multivariate signal in the Fourier domain to reduce confounds from time-correlated noise, thus improving the suitability of the GLM for subject level analysis. Despite these advances, however, the GLM can only \textsl{confirm} hypothesized nonlinear models of function, not \textsl{discover} them.  

Group-level inferences from fMRI have also been performed using linear ICA (independent component analysis), as described in \cite{calhoun2001}. Though ICA and the GLM can be used in conjunction, for example in \cite{liu2010} to investigate the neural effects of stimulation of a particular acupoint, ICA is particularly useful in circumstances that preclude the use of the GLM, such as the analysis of resting-state data, for which there is no task or stimulus regressor.  Covarying networks have been suggested by ICA of resting-state fMRI  in \cite{smith2009}, and functional, hierarchical classification of these networks has been automated through HCA (hierarchical cluster analysis) of aggregated experimental metadata in \cite{laird2011}. However, it was determined early on, for example in \cite{mckeown1998}, that nonlinear interactions within the brain need to be addressed in order to properly determine functional architecture. 

Although ICA algorithms that employ nonlinear mixing functions exist, severe restrictions on those functions are required to avoid non-uniqueness of solutions, as explained in \cite{hyvarinen1999}. Due to this failing, other methodologies have been employed in the attempt to account for nonlinearity.  Examples include various forms of nonlinear regression, as in \cite{kruggel2000}, and dynamic causal modelling, as described in \cite{friston2003}.  In each of these, a particular nonlinear form must be posited \textsl{a priori}, and thus the capability to \textsl{discover} previously unknown nonlinear interactions within the brain is diminished.  As a result, a fuller picture of the nature of intra- and inter-network functional connectivity within the brain is missing from the literature. 

Here we introduce a methodology designed to accomplish such a mathematical characterization, provide insight at the group, subject, and ROI levels, and to avoid linear and univariate assumptions.  With some modification, analysis of higher dimensional data is likely attainable, allowing for eventual application at the voxel scale and eliminating the necessity of ROI selection. After standard preprocessing (slice-timing and motion correction, normalization, smoothing, etc.), our procedure consists of ROI selection, inter-ROI \textsl{symbolic regression} (a model-free form of nonlinear regression), accomplished by an evolutionary algorithm called \textsl{genetic programming} (GP; a form of stochastic optimization), and statistical analysis of the resulting models. We demonstrate our technique on a 242-subject collection of resting-state data from the IMAGEN project, though analysis of task or stimulus experiments can be accomplished with little or no modification. The IMAGEN project is described in detail in \cite{schumann2010}.

We organize the paper as follows. In Section \ref{sec:methods}, we discuss the data and selection of ROI, provide some background on GP, and describe the procedural details of NFM by symbolic regression.  In Section \ref{sec:results}, we report results of applying the technique to the IMAGEN data, including statistical and hierarchical visualizations, comparison with previous results for cross-validation, effects of nonlinearity, and an example of group-level variation. We discuss the results and potential applications of the technique in Section \ref{sec:discussion}, and conclude the paper in Section \ref{sec:conclusions}.   

\section{\label{sec:methods}Materials and methods}

In this section, we first briefly describe the source of the data for our study, and then provide the details of ROI selection that allow for comparison with recent work. We then provide some background on the GP algorithm in general and the specific implementation employed here, along with the method by which it is applied to BOLD signal time series extracted from the selected ROI.  Finally, we describe the statistical technique used to interpret the roughly quarter of a million mathematical models that result from the application of GP to all 52 ROI time series extracted from each of the 242 subjects.   

\subsection{Data}

The data investigated here are a subset of the fMRI scans from the IMAGEN study, a European research project with the goal of better understanding teenage psychological and neurobiological development.  The project is longitudinal, and utilizes several forms of high and low-tech experimental protocols including self-report questionnaires, behavioral assessment, interviews, neuroimaging, and blood sampling for genetic analyses.  Each of the 2000 participating adolescents was 14 when entering the study, which itself commenced in late 2007, and data collection continues today.   

More specifically, the data for the present study are 6-minute resting-state fMRI time series of 242 of the adolescent subjects who were asked to keep their eyes open while in the scanner, but were presented with no other task or stimulus. To allow for comparison with previous work, locations of the ROI were chosen based on results from \cite{laird2011}, in which statistical analysis across thousands of previous imaging studies (both stimulus/task-based and resting-state) was used to identify networks of brain regions that tend to activate together, termed ICN (intrinsic connectivity networks). The ICN were determined by ICA, from which $z$-statistic maps were derived.  To select ROI for this study, a $z$-statistic threshold was set for each ICN to determine the number of regions in the network, and ROI were defined as rough spheres with radii of 3 voxels (9mm) and centered at the location of peak $z$-statistic in each region.  

\begin{figure}[htp]
\centering
\includegraphics[trim=0cm 3.5cm 0cm 3cm, clip=true, width=\columnwidth]{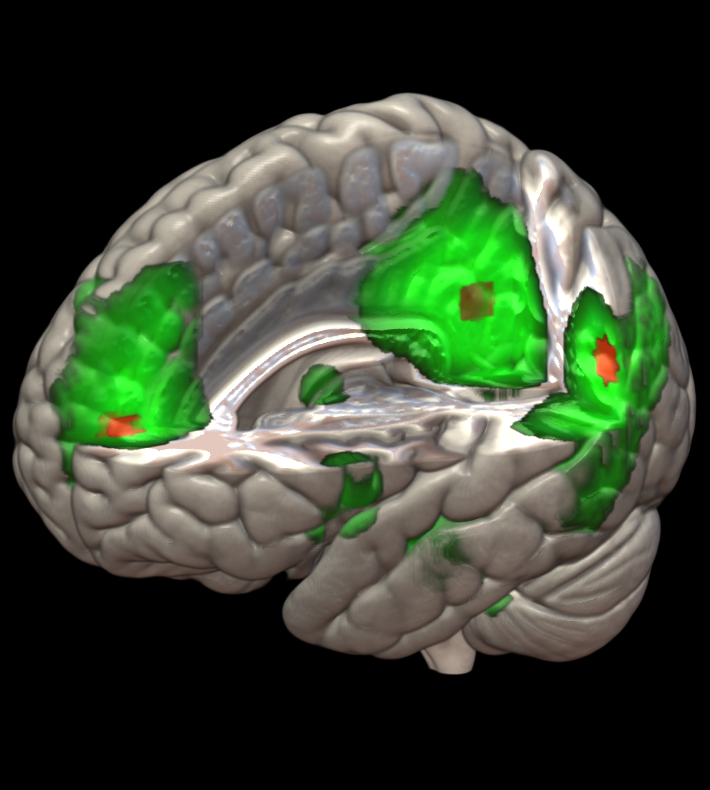}
\caption[ROI Selection]{\label{fig:4roi}ROI Selection. (Red) ROI from within the default mode network, with radii of 3 voxels and centers corresponding to the highest $z$-statistics (green) in each region as determined in \cite{laird2011}.}
\end{figure}

\begin{figure}[ht]
\centering
\includegraphics[width=\columnwidth]{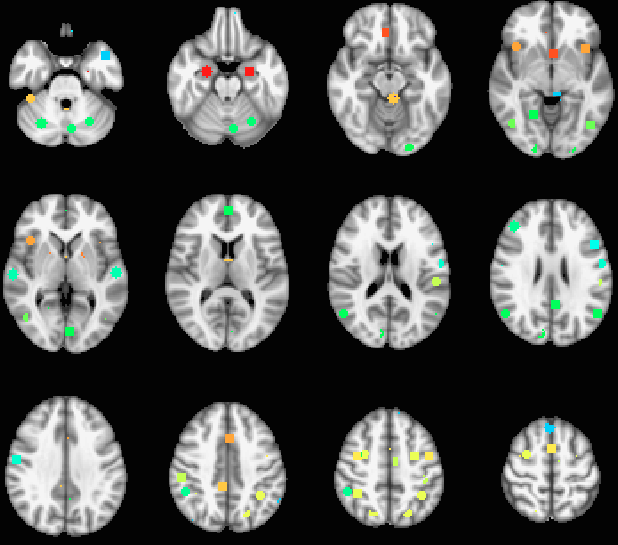}
\caption[Visualization of ROI]{\label{fig:allroi}Visualization of ROI. Axial cross sections showing many of the ROI derived from the ICN in \cite{laird2011}.}
\end{figure}

We provide a cut-out illustrating ROI selection for the default mode network (ICN 13) in Figure \ref{fig:4roi}, and Figure \ref{fig:allroi} contains axial cross sections showing many of the ROI derived from the 18 non-artifactual ICN in Laird et al. (2011).  In \ref{sec:ROI}, Table \ref{tab:ROI} we list all 52 ROI by number, give their anatomical names, indicate the ICN from within which they were defined, and provide visual representations of their locations within the brain. 

Subsequent to ROI definition, a gray matter mask was applied to assure that only appropriate voxels were contained within each ROI.  In some cases this resulted in a considerable reduction of ROI voxels, but the majority maintained the full complement of about 100 voxels.  For each of the 242 subjects, time series were extracted from each of the 52 resulting ROI by averaging the BOLD signal over all voxels within the ROI.  These time series then form the input to the GP algorithm.   

\subsection{Genetic programming}

\begin{figure*}[ht]
\centering
\includegraphics[width=\textwidth]{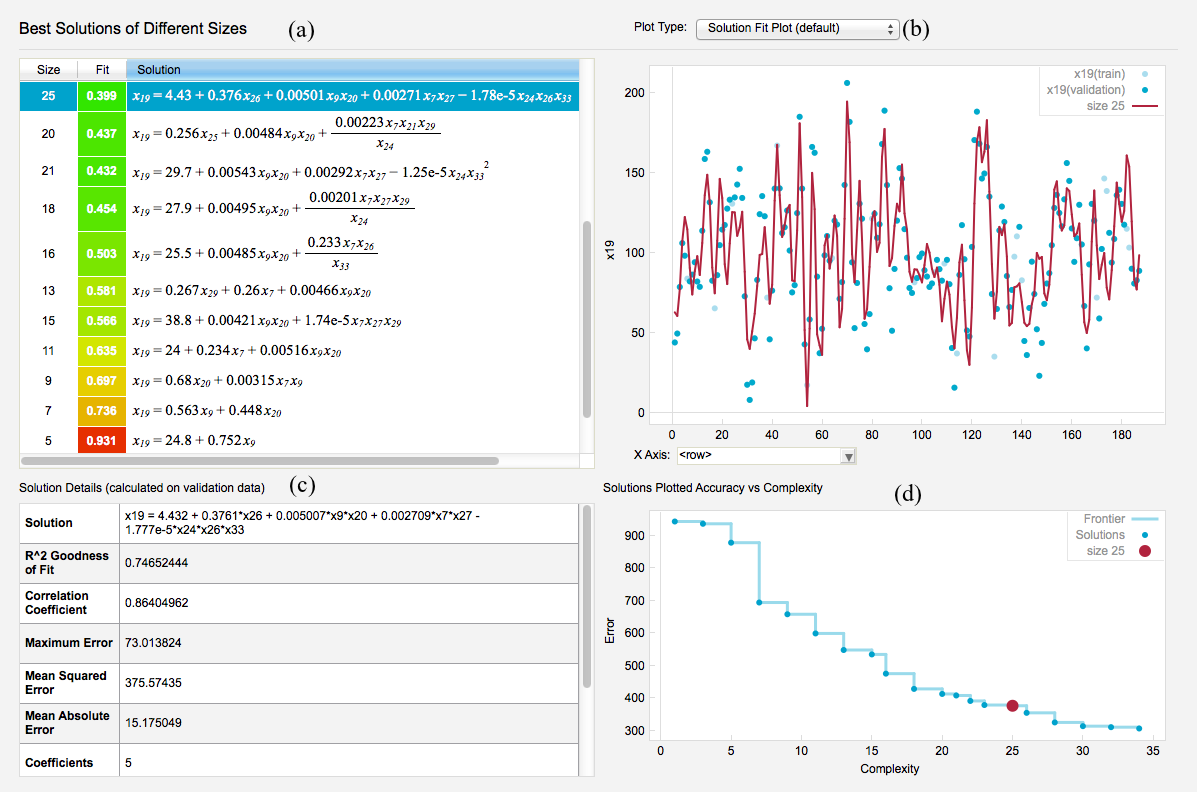}
%\vspace{-.3in}
\caption[Screen shot of the GP package Eureqa]{\label{fig:eqscr}Screen shot of the GP package Eureqa during a search for models of the activity in ROI $19$ in a single subject, as a function of activity in the other 51 regions.  (a) The current set of models along the Pareto front of accuracy vs.\ parsimony, shown in (d) where each point represents a model and the red point represents the highlighted model. (b) Data from ROI $19$ (points) over the 6-minute time series for this subject ($x$-axis in scans, 2 seconds each). The highlighted model is shown in red, and statistics for this model's fit appear in (c).}
\end{figure*}

GP is a biologically inspired, population-based machine learning algorithm. It is most commonly employed for symbolic regression: the algorithm searches for models explaining some quantity of interest (e.g., average BOLD signal from an ROI in the brain) as a function of some other possibly related observable quantities, statistics, or summary data (e.g., BOLD signals from other ROI). The algorithm proceeds by evolving the functional forms of a population of potential models, which are initially constructed at random from user-specified mathematical building blocks (available variables, arithmetic functions, parameter constants, etc.).   In brief, the models that better explain the data produce more offspring, leading to a gradual reduction of error within the population.  We show a representative set of models produced by this approach in Figure \ref{fig:eqscr}(a). A key advantage of the technique is that no assumption (e.g., linearity) is imposed on the form of solutions, other than the choice of building blocks from which they can be made (we use arithmetic operations in the present work). 

Typically, some measure of error (e.g., root mean square error) constitutes a model's explanatory fitness, and some measure of its size (e.g., number of operators, constants, and variables in the equation) represents its parsimony. The next generation of potential models is obtained by mutation (e.g., a single change of variable or operator) and recombination (i.e., swapping of function components between models) of the current set of \textsl{non-dominated} solutions: those models for which no simpler model in the population has less error. This set of non-dominated models is said to approach the ideal \textsl{Pareto front} of fitness versus parsimony as the population evolves. An important aspect of GP is that the result of a single search is this entire set of potential models, providing a trove of information for statistical analysis.  Figure \ref{fig:eqscr} is a screenshot of the off-the-shelf GP package Eureqa from \cite{schmidt2009} performing a search (Eureqa version 0.97 Beta was used to generate the results reported in this study).

To apply GP to the fMRI data, for each of the 242 subjects we extract a single BOLD signal time series from each of the 52 selected ROI by averaging over the voxels within that ROI.  Then the GP algorithm is run 52 times, one for each ROI, using all other ROI as potential explanatory variables.  Note that the algorithm has no knowledge of the hypothesized networks from which these regions were chosen.  

We describe the computational expense of the algorithm in terms of core-hours, i.e., the number of hours required for a single processor core to perform the necessary computation.  Specifically, twelve core-hours of search were performed for each region, amounting to 624 core-hours per subject, and over 17 total core-years of computation were required for the population of 242 subjects.  This yielded roughly 12 thousand Pareto fronts comprised of a quarter million models for statistical analysis.  

The results of this analysis characterize the entire population of 242 subjects.  Alternatively, results can be aggregated over phenotypic groups to produce group-level characterizations, or many GP searches can be run for a single individual to produce a subject-level characterization.  We report results of population- and group-level analyses in Section \ref{sec:results}, and discuss an example subject-level analysis in \ref{sec:indvar}.  

\begin{figure*}[ht]
     \centering
     \subfigure[Interaction map]{
          \label{fig:imap}
          \includegraphics[width=.46\textwidth]{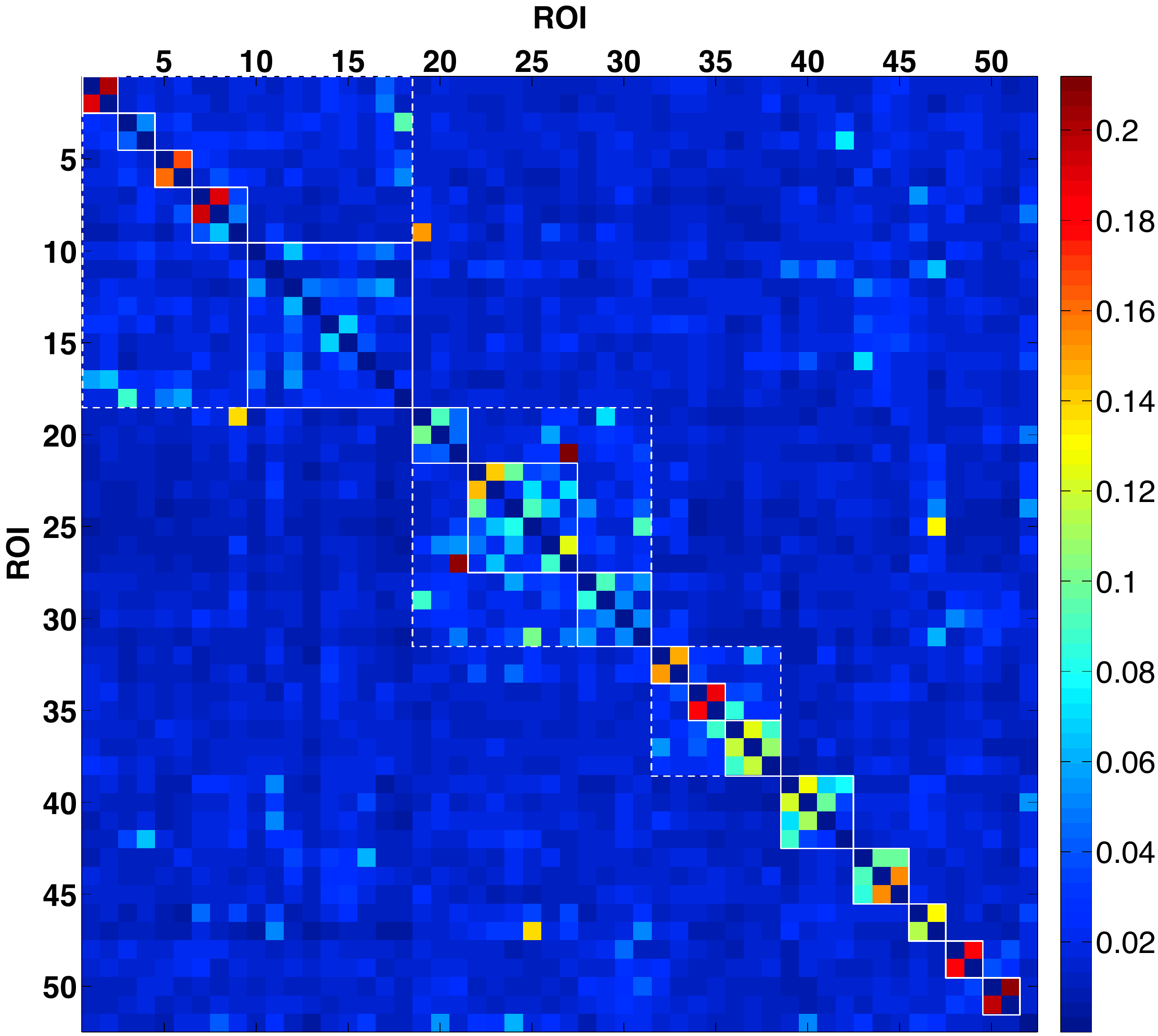}} 
    \hspace{.1in}
     \subfigure[Subsampling RSD]{
          \label{fig:imaprsd} 
          \includegraphics[width=.46\textwidth]{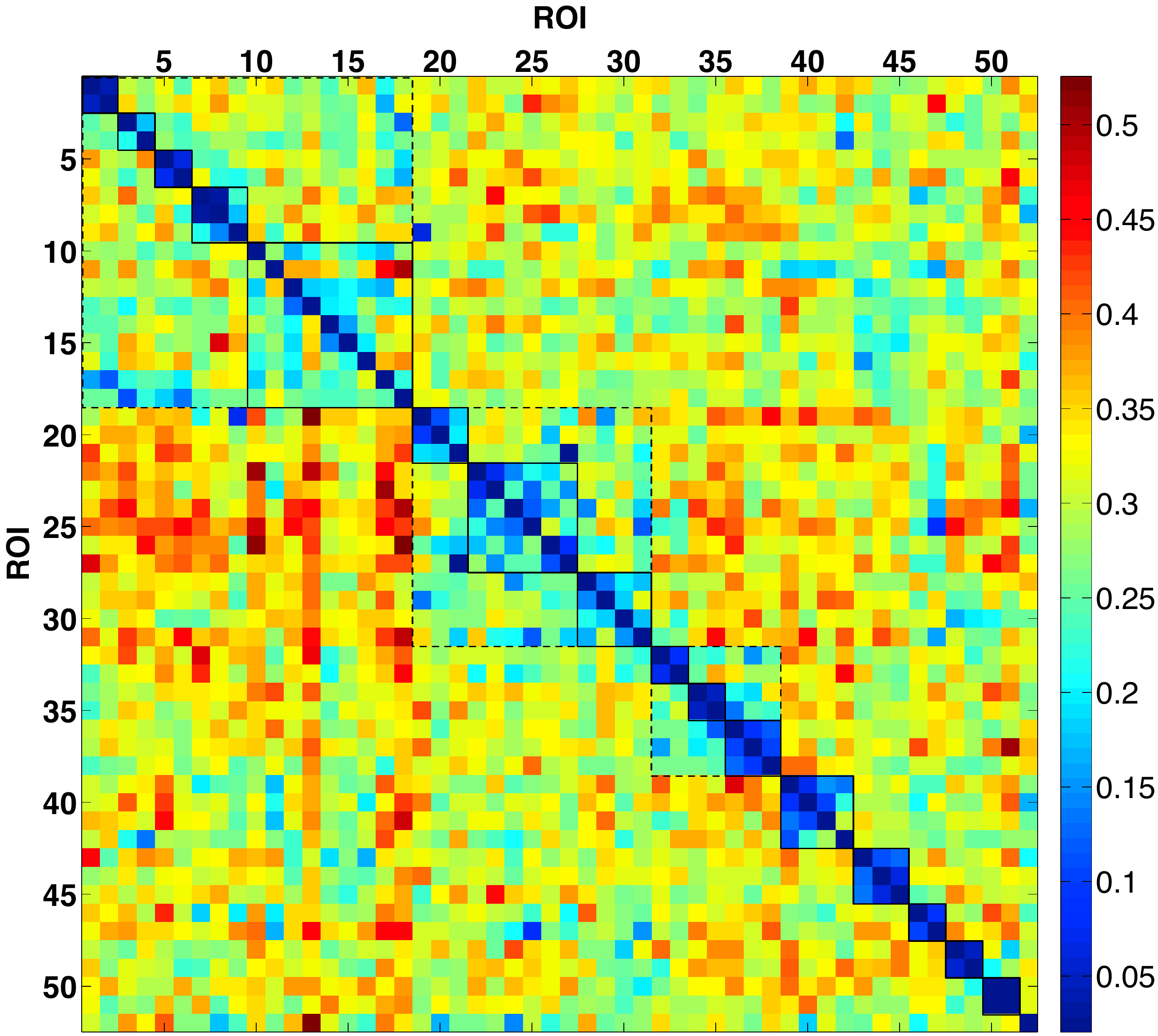}}
          %\vspace{-.1in}
     \caption[Functional interaction map]{\label{fig:imapstats}Functional interaction map. (a) Interaction map across all 242 subjects, and (b) map of RSD (relative standard deviation) of the interaction rates over 100 subsamples with 100 randomly selected subjects each.  Solid outlines indicate ICN and dashed outlines indicate functional groupings of ICN from \cite{laird2011}.}  
\end{figure*}

\subsection{Analysis}

The output of the GP algorithm poses a challenge for interpretation.  Here we present a coarse statistical analysis of this rich mathematical characterization.  For each ROI, we count the number of models for that ROI, across the Pareto fronts for all 242 subjects, that have a particular (other) region on the right-hand side of the equation.  We compute this count for each of the other 51 regions. 

For example, consider the GP search for models of ROI $19$ within a single subject illustrated in Figure \ref{fig:eqscr}.  Upon completion, all 20 of the models along the Pareto front for this subject had at least one term containing ROI $9$, and 17 models had terms containing ROI $20$.  In the subject pool as a whole, the total counts are 2990 and 1984, respectively.  Specifically, of the roughly 5000 models for ROI $19$ across all subjects, about 60\% have terms containing ROI $9$, and about 40\% have terms containing ROI $20$.  Note that these frequencies are not properly normalized, because most models contain several ROI. Thus we normalize by the sum of the counts for all ROI. In the case of ROI $19$, this sum is 22016.   

The result is a vector for each ROI that describes, in a statistical sense, its relative dependence on each of the other regions.  We interpret this vector as a distribution of likely interaction, and define the computed values to be relative \textsl{interaction rates} (IR).  Note that both linear and nonlinear interactions, as well as weakly and strongly weighted basis functions, are counted equally.  We form an interaction map by stacking these IR row vectors to visualize interaction across all 52 ROI, shown in Figure \ref{fig:imap}.  The value in row 19 column 9, for example, is $2990/22016\approx0.136$, depicted as a yellow square.  

Note that the IR map is not symmetric by construction (though it appears nearly so), and indeed the value in row 9, column 19 is $0.148\ne0.136$.  We interpret a row of the IR map as a distribution of relative \textsl{dependence} of the corresponding ROI on each of the other regions. We interpret a column, on the other hand, as a measure of the \textsl{influence} of the corresponding ROI on each of the other regions. By averaging the IR map with its transpose, we produce a symmetric, \textsl{overall} IR map (not shown) that can be used in hierarchical analysis. We examine the interaction map, and provide results of hierarchical analysis, in the next section.

\section{\label{sec:results}Results}

Figure \ref{fig:imap} shows the interaction map generated by the normalized frequency analysis of the NFM procedure, summarizing ROI interaction across all 242 subjects. To test the robustness of the computed interaction map, we form 100 random subsamples (with replacement) from the pool of 242 subjects, each with 100 subjects.  For each sample, we perform the same counting procedure to produce the interaction map corresponding to that sample.  A heat map of relative standard deviation (RSD) of IR over the 100 subsamples is shown in Figure \ref{fig:imaprsd}. 

The strong block-diagonal structure of the interaction map corresponds directly to the grouping of ROI into ICN.  For example, regions 39-42, which form a partial block in the figure, are the four ROI that make up the default mode network (ICN 13) in \cite{laird2011}. Robustness (across subjects) of intra-network interaction is supported by the matching block-diagonal structure of low subsampling RSD (mean intra-network RSD $< 20\%$), for all but ICN 2 (ROI 3-4) and ICN 5 (ROI 10-18).  In addition to the strong primary block-diagonal structure, there is a secondary structure of lighter blocks that group ICN together.  For example, regions 32-38 are composed of the strong blocks 32-33, 34-35, and 36-38 (corresponding to ICN 10, 11 and 12 respectively).  There is a lighter block structure that suggests interaction among these three ICN which are, in fact, \textsl{together} responsible for visual processing. The secondary structure visible for regions 19-31 is comprised of ICN 6-8, which perform motor and visuospatial tasks.  Each of these examples shows a matching secondary structure of moderate subsampling RSD (mean inter-network RSD $< 30\%$), indicating fairly robust inter-network interaction as well.

%\begin{figure*}[ht]
%\centering
%\includegraphics[width=\textwidth]{dendro.pdf}
%\caption[HCA of interaction among ROI]{\label{fig:dendro}Hierarchical cluster analysis (HCA) of interaction among ROI.  The distance between regions is taken to be the reciprocal of the IR between them, so that regions with an IR of 0.2 between them (ROI 1,2) have an interaction distance of 5. The organization of ROI into networks, and clustering of those networks into functional groups described in \cite{laird2011} are each indicated. }
%\end{figure*}

\subsection{Hierarchical analysis}

To further illustrate and clarify the hierarchical organization suggested by the interaction map, we generate the dendrogram in the top of Figure \ref{fig:dendrolinnonlin} by HCA (hierarchical cluster analysis, implemented in MATLAB with the nearest distance algorithm), using the reciprocal of the overall IR between each pair of ROI as the distance between them.  For example, ROI 1 and 2 have an approximate overall IR of 0.2, and thus the distance between them is 5. We emphasize that the organization of ROI into networks, and clustering of those networks into functional groups described in \cite{laird2011}, are both captured by NFM. Some examples:     

\begin{figure*}[htp]
\centering
\subfigure{\includegraphics[width=\textwidth]{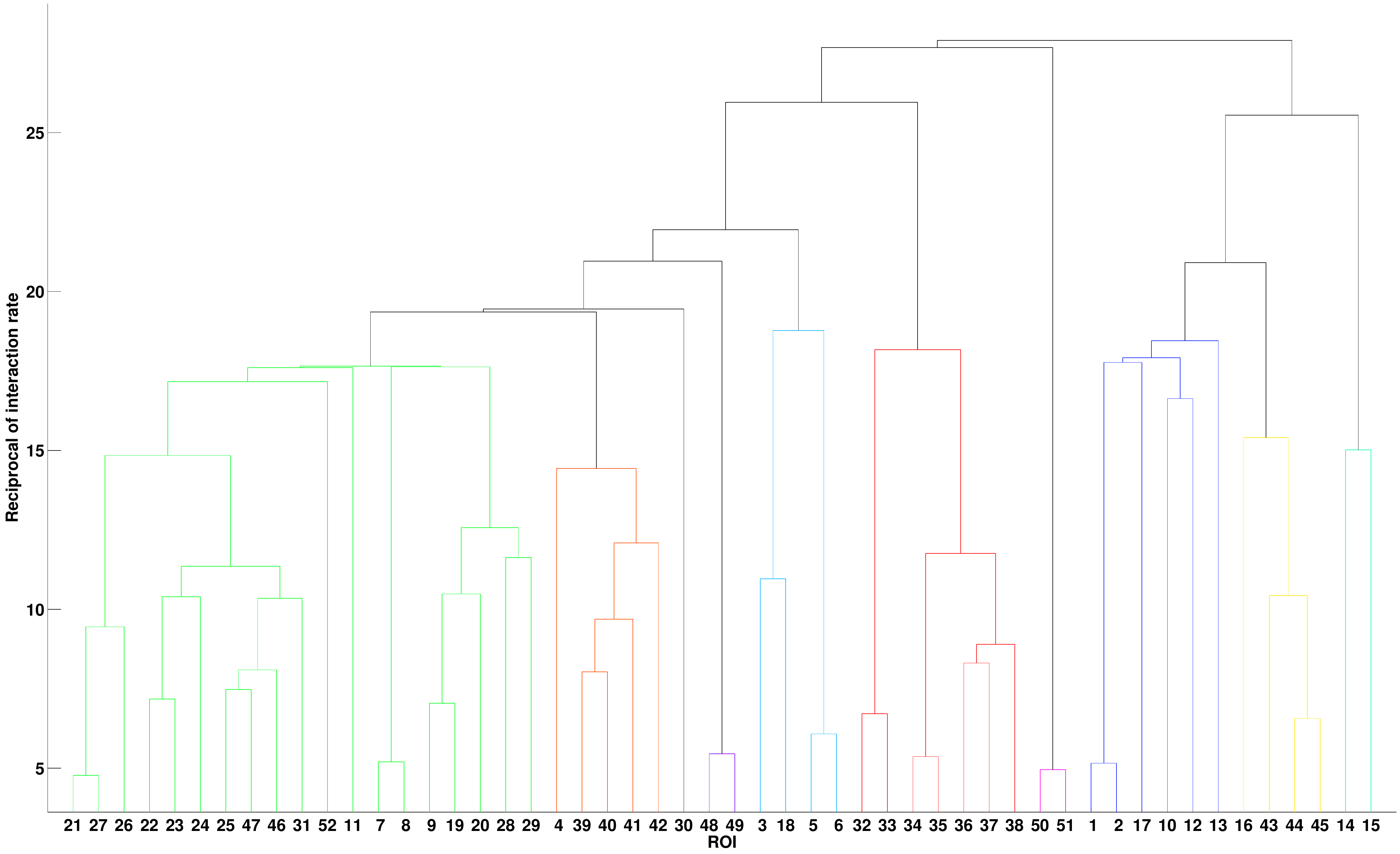}}\\
\subfigure{\includegraphics[width=\textwidth]{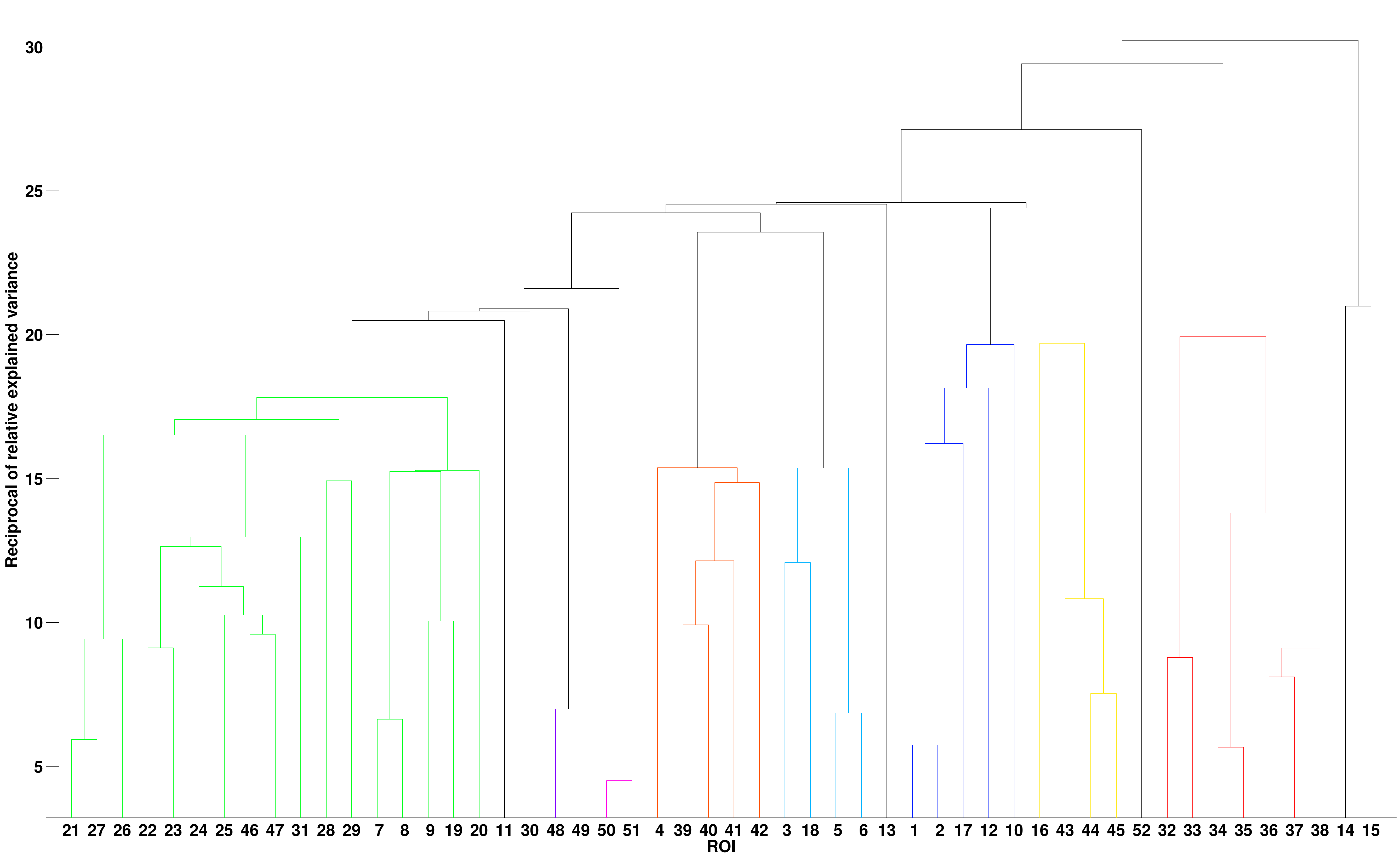}}
\caption[HCA of interaction among ROI]{\label{fig:dendrolinnonlin}Hierarchical cluster analysis (HCA) of interaction among ROI, generated with NFM (top) and correlation analysis (bottom).}
\end{figure*}

\begin{itemize}
\item The red group forms the visual cluster. ROI 32 and 33, the lateral occipital cortices, form one network (ICN 10), while ROI 34-35, the occipital poles, and ROI 36-38, the lingual gyrus, right cuneus and right fusiform gyrus, respectively, form two other networks (ICN 11 and 12) from within the visual cluster.
\item Regions 39-42 (the orange group) form ICN 13, the default mode network, and interact with ROI 4, the ventromedial prefrontal cortex, from ICN 2.
\item The green group to the far left includes all but one of the ROI from the motor and visuospatial complex. Interaction of this complex with the middle cingulate cortex (mCC, ROI 9) and the network composed of ROI 46 and 47, thought to be responsible for multiple cognitive processes such as attention and inhibition, is indicated as well, suggesting that this interaction was common among many of the subjects.   
\item ICN 1 (ROI 1,2), 3 (ROI 5,6), the first two regions from ICN 4 (ROI 7-9), ICN 14 (ROI 43-45), 16 (ROI 48,49), and 17 (ROI 50,51) are also indicated.  
\item Many of the regions from ICN 5 (ROI 10-18) interact with ICN 1 (ROI 1-2), and also form a loose interaction group with ICN 14 (ROI 43-45), the cerebellum, the most robust connection of which appears to be between ROI 16 and 43.
%\item We note the apparent lack of interaction among networks from the emotional/interoceptive functional group, ICN 1-5.  Each network appears to either be isolated, or interacting outside the group (in the case of ICN 5 (ROI 10-18), in particular). 
\end{itemize}
\noindent
The robustness of each of the interactions discussed in this list is supported by low interaction rate subsampling RSD, shown in Figure \ref{fig:imaprsd}.

\subsection{\label{sec:linvnonlin}Impact of nonlinearity}

In this section we demonstrate that the NFM procedure both captures the hierarchical structure of ROI interaction indicated by linear analyses, \textsl{and} reveals nonlinear interactions not discoverable by such methods.  To accomplish this, we compare the population-level hierarchy generated by NFM with the results of an analogous linear procedure involving pairwise correlation analysis. Furthermore, we validate nonlinear relationships suggested by NFM in a stepwise multiple regression, and an elastic net regularized regression, the results of which we describe at the end of this section.

\subsubsection{Comparison with correlation analysis}

For each of the 242 subjects, we compute the correlation matrix for the 52 ROI time series.  Squaring the elements of the correlation matrix and normalizing each row (after setting the diagonal to zero) provides the relative explained variance (relative $R^2$) of the ROI corresponding to that row by each of the other 51 ROI. The average of the 242 normalized subject matrices is interpreted as the linear version of the population-level IR map generated by NFM. As with IR, the reciprocal of relative explained variance can be considered a distance between ROI (higher relative $R^2$ means closer). The resulting hierarchy generated by HCA is shown in the bottom of Figure \ref{fig:dendrolinnonlin}.

As expected, much of the large-scale structure revealed by NFM is also indicated by the linear correlation analysis.  The similarity of the generated hierarchies supports the validity of the models discovered by GP (i.e., the algorithm is not excessively overfitting the data), and the subtle differences between them suggest potentially interesting interactions that are missed if linearity is assumed. In the following we investigate one of these differences.

\begin{figure}[htp]
\centering
\includegraphics[trim=0cm 4cm 0cm 4cm, clip=true, width=\columnwidth]{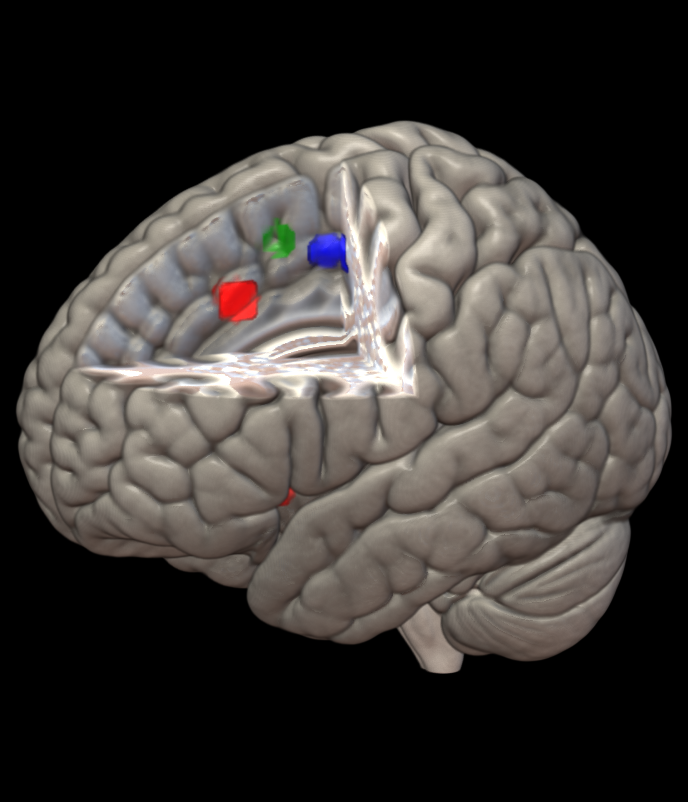}
\caption[NFM reveals a nonlinear interaction]{\label{fig:ac-pm-cs} NFM reveals a nonlinear interaction among these three ROI: mCC (red), posterior dmPFC (green), and PL in the primary motor cortex (blue). }
\end{figure}

Interaction of the mCC (ROI 9) with the motor visuospatial complex is evident in both hierarchies.  However, in the linear analysis it appears more closely connected with its own ICN (ROI 7,8, the bilateral anterior insula), and only with the posterior dorsomedial prefrontal cortex (dmPFC, ROI 19) from ICN 6. In contrast, NFM reveals that activity in the mCC is related to more components of the motor system.  The nonlinear models generated by GP show a strong connection between the mCC, posterior dmPFC, and the paracentral lobule (PL) of the primary motor cortex (ROI 29 from ICN 8) shown in red, green, and blue, respectively, in Figure \ref{fig:ac-pm-cs}.  Specifically, about 20\% of all of the models generated for the activity in the posterior dmPFC, across all subjects and levels of complexity, contain \textsl{both} the mCC and the PL as explanatory variables.

For many of these models, the mCC and PL only show up as linear terms, so it is reasonable to wonder why the correlation analysis did not pick up this interaction.  The vast majority of models containing mCC and PL in only linear terms \textsl{also} contain nonlinear terms in other ROI.  It is the posterior dmPFC \textsl{along with} these nonlinear terms that is correlated with the mCC and PL. Thus the interaction is hidden from linear analyses.  Furthermore, many of the models \textsl{do} contain nonlinear terms involving the mCC and PL.  In fact, the product of the activity in these two regions shows up in 78 models for the posterior dmPFC across 21 different subjects, and it is always additive.  This term is involved in models across the spectrum of complexity, including instances where it is the \textsl{only} term.  

% The following probably goes without saying:
%We can not directly infer causality from this term with our data, but NFM revealed the relationship.  

\subsubsection{Validation of nonlinear terms}

To validate first order nonlinearity (pairwise product and quotient terms, as well as reciprocals) suggested by NFM, we first randomly assign 100 subjects to a training group, and 100 different subjects to a testing group.  NFM results are aggregated over the training group to produce an IR map and hierarchy (not shown) summarizing ROI interaction within the training group as a whole.  The roughly 2000 specific models generated by NFM for each region (approximately 20 models per training subject) are then used to inform the modeling of ROI activity in that region within the testing group, by stepwise regression. 

For each ROI and each testing subject, we first perform a standard stepwise linear regression using the other 51 ROI as regressors.  We then perform a stepwise regression including all first order nonlinear terms suggested by NFM over the training group in addition to the 51 linear regressors.  Statistics of the linear and nonlinear models are compared to determine the effect of including these first order terms. To illustrate, we describe results of the validation procedure for the posterior dmPFC (ROI 19) here. 

\begin{figure}[ht]
\centering
\includegraphics[width=\columnwidth]{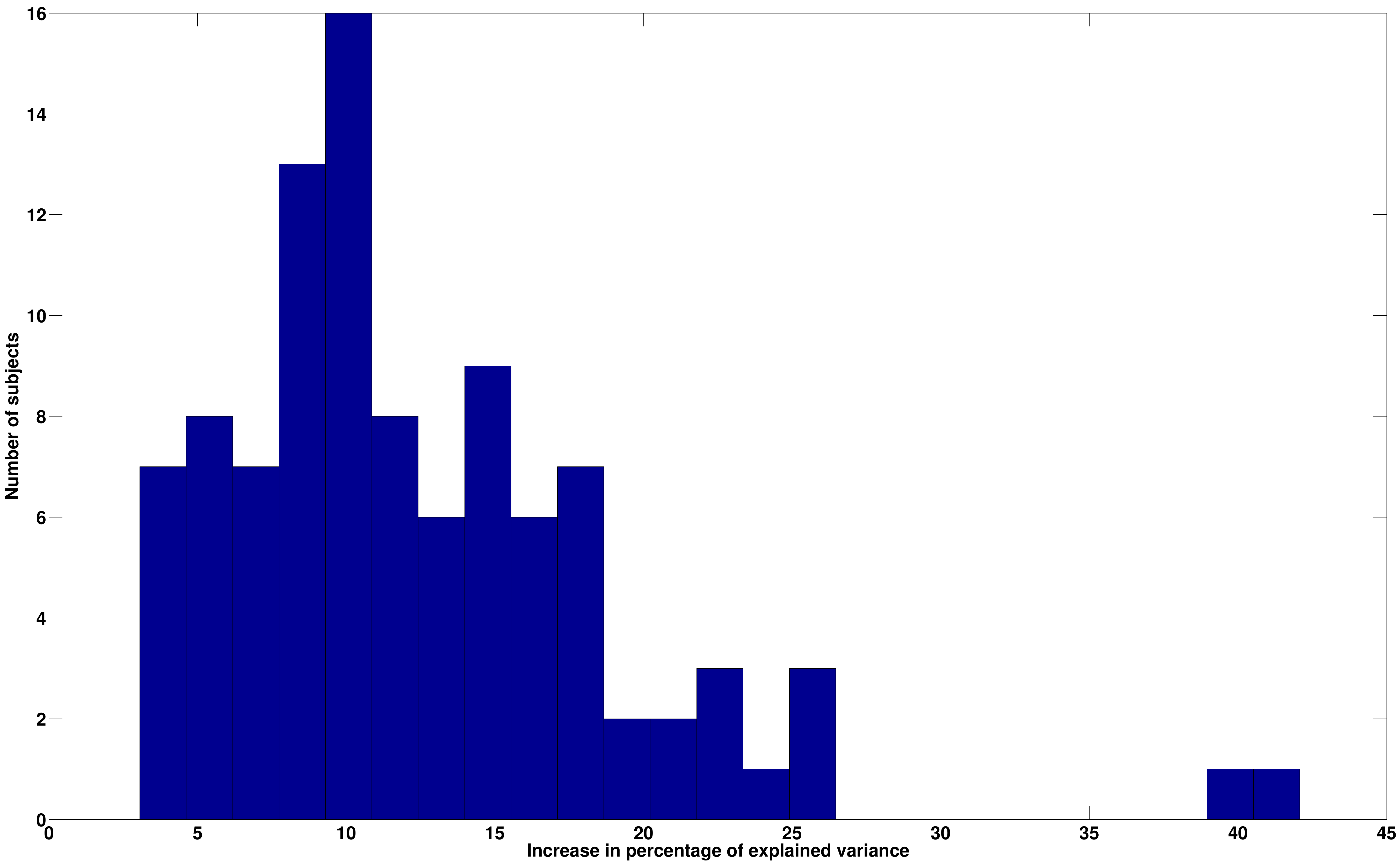}
\caption[Histogram of increase in explained variance]{\label{fig:rsqboosthist}Histogram of increase in explained variance. The inclusion of first order nonlinear terms suggested by NFM over the training group, in a stepwise regression analysis for the posterior dmPFC in the testing group, increases the percentage of explained variance for \textsl{every} test subject, with a mean increase of 12.5\% and maximum increase of 42\%. }
\end{figure}

We show a histogram of increase in the percentage of explained variance for the nonlinear versus linear regression models for the posterior dmPFC in Figure \ref{fig:rsqboosthist}. The inclusion of first order nonlinear terms suggested by NFM over the training group increases the percentage of explained variance for \textsl{every} test subject, with a mean increase of 12.5\% and maximum increase of 42\%. The nonlinear models contain more terms (mean 46, compared with mean 19 for linear models), so a potential concern is that the increase in $R^2$ might simply be a result of the additional degrees of freedom.  However, for each test subject the nonlinear model $F$-statistic is also greater than that of the linear model (mean increase of 83.5, maximum increase of 1000), and comparisons of adjusted $R^2$, which account for differences in degrees of freedom, show only slightly smaller increases for all test subjects.  This suggests that the increase in explained variance is due to explanatory power of the nonlinear terms, and not simply the additional degrees of freedom in the nonlinear models. 

To further support the validity of the nonlinear terms suggested by NFM, we apply a similar testing approach using a machine learning algorithm called elastic net regularized regression.  In contrast to stepwise regression, regularization allows for the inclusion of highly correlated explanatory variables, while simultaneously discounting regressors with very small coefficients.  Regularized models can have more explanatory power or fewer terms (or both), with respect to those from stepwise regression.  We see each of these scenarios in the present case.  Using the same training and testing groups, and modeling the same ROI (the posterior dmPFC), elastic net regularization produces linear models with an average of 45 terms that explain roughly the same amount of variance (on average) as the nonlinear models generated with stepwise regression, improving upon the explanatory power of their stepwise counterparts.  However, the regularized \textsl{nonlinear} models provide that same explanatory power with a mean of only 26 terms.  Furthermore, the regularized nonlinear model is preferable to the regularized linear model, as determined by the Akaike information criterion, \textsl{for every single test subject}.              

\subsection{Group-level variation}

\begin{figure*}[htp]
\centering
\subfigure{\includegraphics[width=\textwidth]{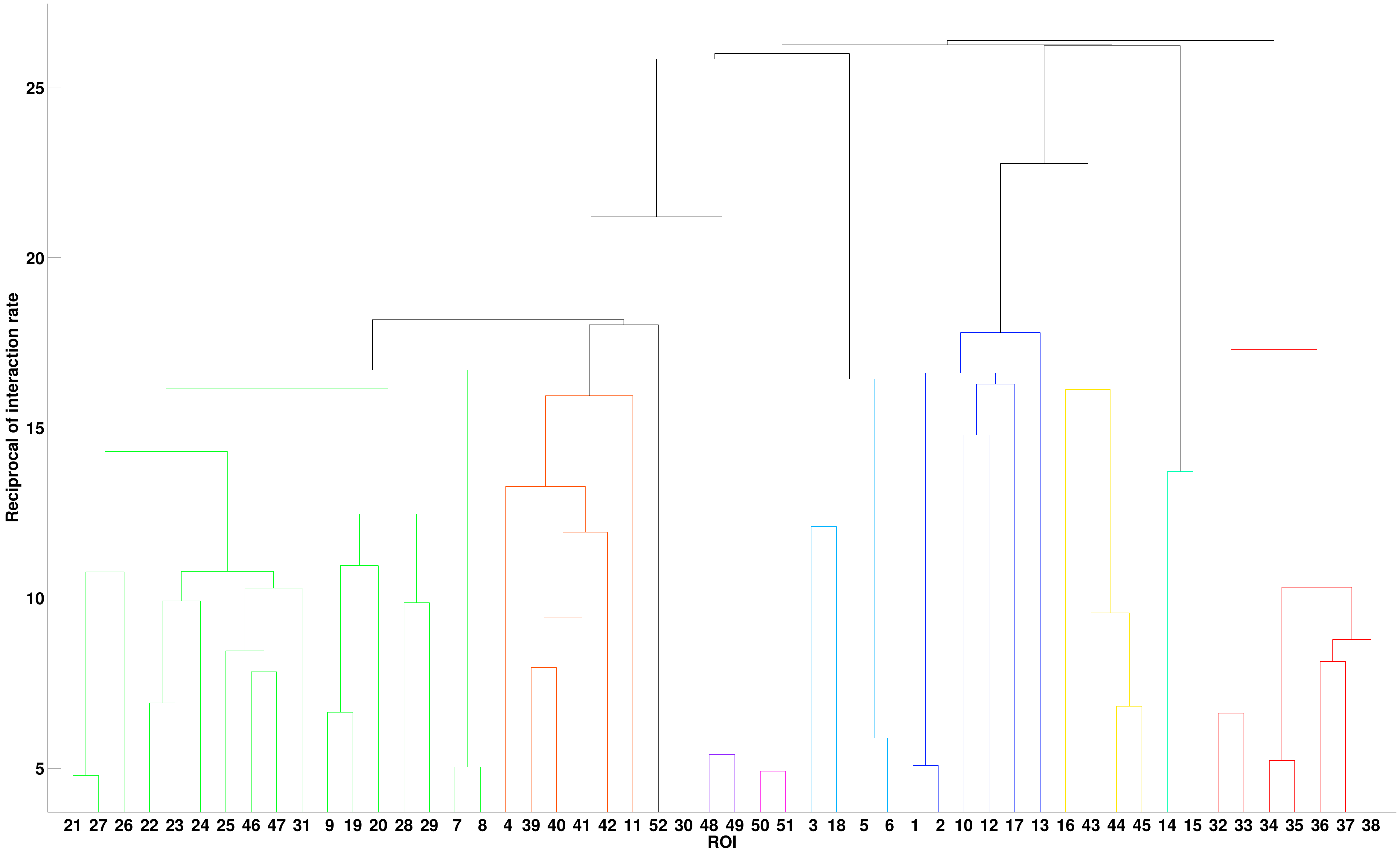}}\\
\subfigure{\includegraphics[width=\textwidth]{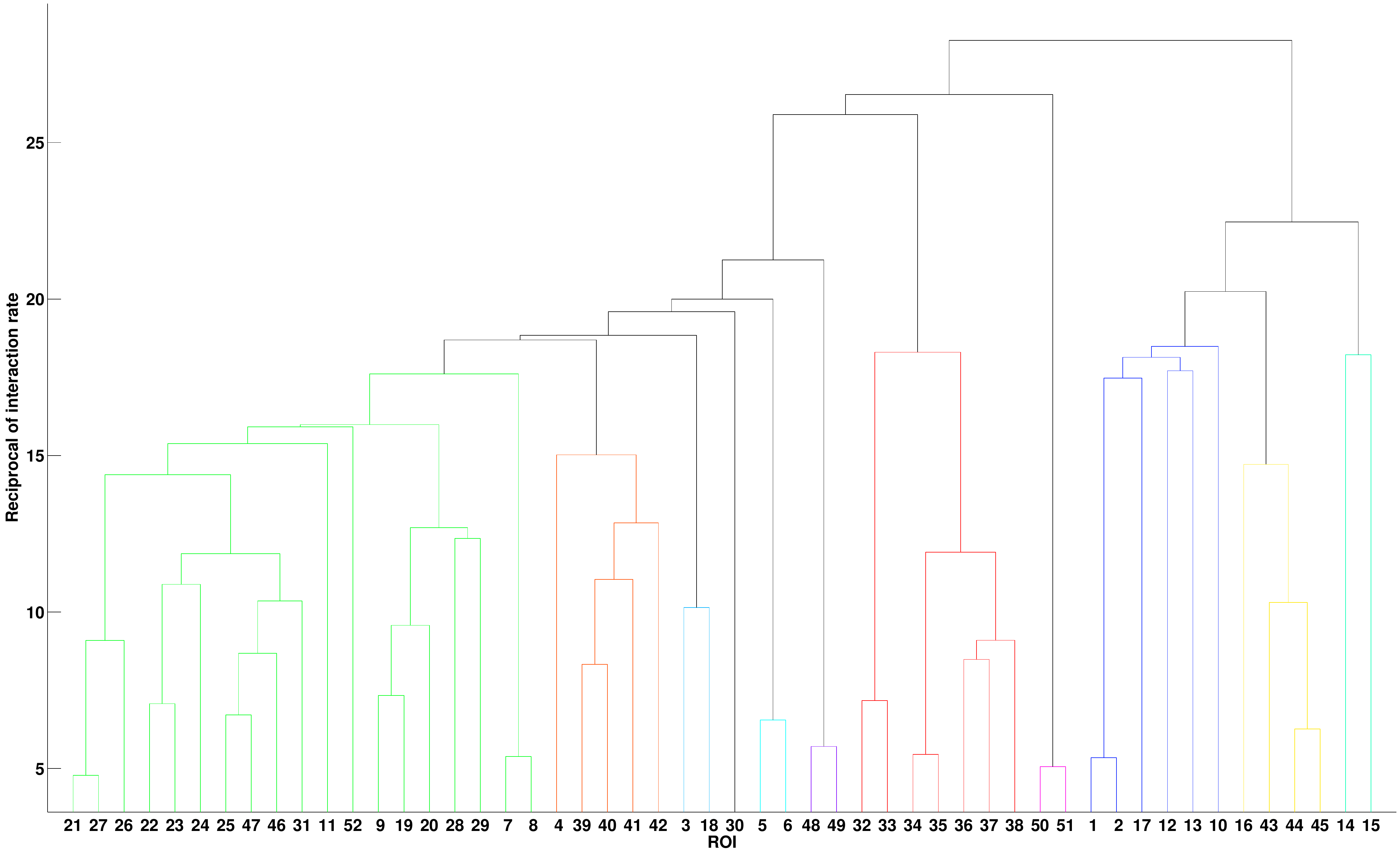}}
\caption[Hierarchies for groups with high and low alcohol consumption rates]{\label{fig:dendroalc}Hierarchies for groups with high (top) and low (bottom) alcohol consumption rates, defined by two or more lifetime drinks and one or fewer lifetime drinks, respectively.}
\end{figure*}

Variation among individuals (illustrated in \ref{sec:indvar}) suggests that statistics of interaction rates among ROI may differ between phenotypic groups.  The hierarchical organization of ROI induced by IR might illuminate, in such cases, variation in functional dynamics associated with demographic, behavioral, or genetic characteristics. An example illustrating this potential is provided by the contrast between drinking (D) and non-drinking (ND) adolescents from the IMAGEN dataset. In Figure \ref{fig:dendroalc}, we show hierarchies for the top and bottom 100 subjects in terms of lifetime drinking score, determined by self-report questionnaire, corresponding to those who have had 2 or more lifetime drinks, and those who have had 1 or fewer, respectively. The two hierarchies are similar to one another (and comparable to the population level hierarchy), but subtle differences between them suggest group-differentiating factors. 

\begin{itemize}
\item The ROI pair 3,18, the subgenual anterior cingulate cortex (ACC) and fornix body, respectively, are coupled in both the D and ND groups.  However, their arrangement in the hierarchies is different, as we'll describe in a moment, resulting from the following two distinguishing interaction rates.
\item For the ND group, there is a 22\% lower IR between ROI 6, the left globus pallidus, and the fornix body, ROI 18.  We note that this reduced interaction is completely missed by pairwise correlation analysis, (which indicates a slightly reduced interaction among \textsl{drinkers}, see \ref{sec:alclin}), and thus appears to be an entirely nonlinear effect.
\item In contrast, there is a 33\% higher intra-network IR within ICN 2, comprised of ROI 3-4, the subgenual ACC and the ventromedial prefrontal cortex (vmPFC), respectively, among non-drinkers.  Though this difference is also indicated by correlation analysis, only about half of the effect is captured (a 16\% elevation).
\item These two differences in interaction cooperate to shuffle the hierarchical arrangement of ROI in the D versus ND group. The subgenual ACC and fornix body are most closely associated with the default mode network in non-drinkers, through the vmPFC. Among drinkers, in contrast, they are grouped directly with the bilateral globus pallidus of ICN 3 (ROI 5-6). In other words, in drinkers there is a tighter coupling among ROI most strongly linked to reward and thirst tasks as reported in \cite{laird2011}.  The relevant ROI are shown in Figure \ref{fig:diffroi}.

\begin{figure}[htp]
\centering
\includegraphics[trim=1cm 2cm 1cm 1cm, clip=true, width=\columnwidth]{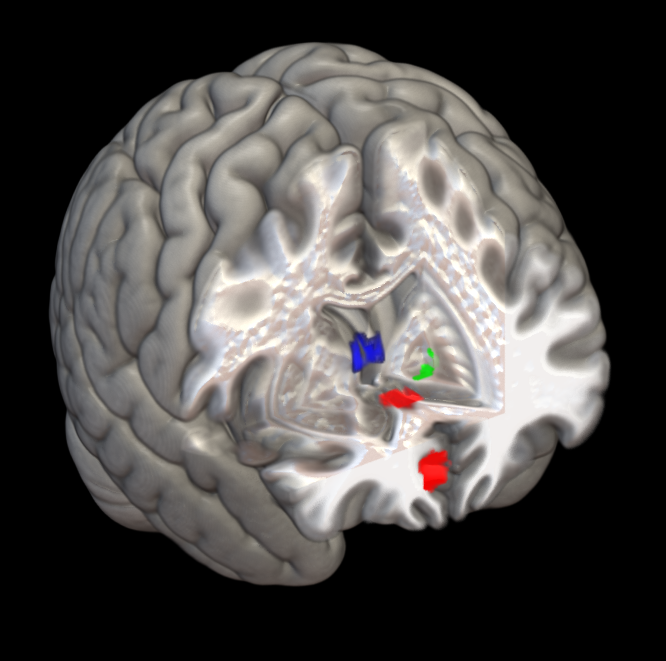}
\caption[Interaction between the sgACC and vmPFC is lower among drinkers]{\label{fig:diffroi} Interaction between the subgenual ACC (top red) and vmPFC (bottom red) is lower among drinkers, who also show elevated interaction between the left globus pallidus (green) and fornix body (blue), an apparently nonlinear effect.}
\end{figure}

\item The largest single difference between the D and ND groups is a 74\% elevated IR between the right angular gyrus (ROI 41 in the default mode network) and ROI 11, the posterior cingulate cortex, among drinkers.  These ROI are shown in red and green, respectively, in Figure \ref{fig:ag-pc}.  About half of this effect is captured by correlation analysis.

\begin{figure}[htp]
\centering
\includegraphics[trim=0cm 4cm 0cm 4cm, clip=true, width=\columnwidth]{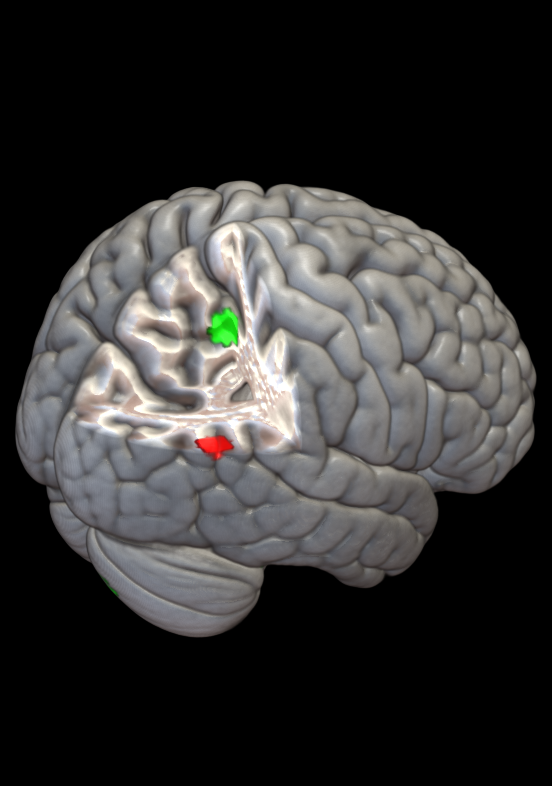}
\caption[Interaction between rAG and PCC is 74\% higher among drinkers]{\label{fig:ag-pc} Interaction between the right angular gyrus from the default mode network (red), and posterior cingulate cortex (green) is 74\% higher among drinkers.}
\end{figure}

%\item Finally, we note that none of these differences in IR are sensitive to adjustments in the definition of the D and ND groups.  The specific results presented come from taking the bottom 100 and top 100 subjects, in terms of the ESPAD 8a lifetime drinking score, as the ND and D groups, respectively.  The result is an ND group that contains adolescents who have had one or fewer lifetime drinks, and a D group with those who have consumed two or more lifetime drinks.  Taking the bottom and top 50 (ND = one or fewer, and D = three or more drinks), or the bottom and top 120 (ND = two or fewer, D = two or more drinks), produces no qualitative change in the results described.
\end{itemize}

\section{\label{sec:discussion}Discussion}

The large extent of ICN reproduction, and their hierarchical organization into functional groups using an entirely different approach than that described in \cite{laird2011}, provides strong evidence for the analytical potential of NFM.  Furthermore, the technique reveals nonlinear interactions that are not discoverable with standard linear techniques, or without prior hypotheses.  Such relationships could provide a new window into brain function, and this highlights the potential of the methodology as a \textsl{hypothesis generator}. Of course proper care must be taken (with regard to independence of observations, etc.) in the ensuing investigations of such data-driven hypotheses.  Nonetheless, hypothesis generation is a powerful tool for scientific exploration, and has been used recently to inform biomedical research, such as in \cite{abedi2012} and \cite{spangler2014}.

In addition to providing insight on its own, the NFM procedure complements other modes of analysis.  A potentially promising extension, especially for a hybrid version capable of voxel level analysis (discussed in \ref{sec:improvements}), would be to use it in conjunction with graph-theoretic analyses such as those described in \cite{bassett2006}, \cite{stam2007}, and \cite{vdh2008}.  The general technique, as detailed in \cite{bullmore2009} and \cite{rubinov2010}, is to compute pairwise correlations among all voxels, set a threshold above which two voxels are considered connected, and calculate various network summary measures (e.g., degree distribution, assortativity, diameter, etc.).  By simply replacing correlations in these networks with interaction rates determined by NFM, the assumption of linearity is left behind.

Finally, it is important to note that the specific forms of the models in the output of the GP algorithm have been analyzed simplistically in the present work.  A major potential benefit of NFM is the insight that might be gained from precisely analyzing these mathematical descriptions of the relationships among ROI in the brain.  Of course, ascribing meaning to any particular one of these models would have to be done cautiously.  However, given the results we describe here, obtained by a coarse treatment, the \textsl{collection} of models determined by GP may offer a number of as yet undiscovered insights.  This seems a potentially fruitful avenue for future theoretical research. 

\section{\label{sec:conclusions}Conclusions}

Results produced in our study suggest that there is potential analytical power in the use of NFM, or some modification thereof, in the neuroimaging domain.  The procedure we investigated here utilizes commercially available, out-of-the-box GP software, and preliminary statistical analysis of its output.  Many improvements and extensions are possible, only some of which we have suggested in this work.  Reproduction of recent results constitutes a measure of cross-validation, and the preliminary results presented demonstrate the unique capability of NFM to discover nonlinear relationships among regions of the brain that hold promise for illuminating differences in brain function between subject groups. Further, the mathematical characterizations we have achieved, which are not limited by linear or univariate assumptions, are ripe for future investigation.   

\section*{Acknowledgements}

This work was supported by DARPA grant FA8650-11-1-7155, and the Vermont Advanced Computing Core, NASA (NNX-08AO96G) at the University of Vermont, which provided High Performance Computing resources. The IMAGEN study receives research funding from the European CommunityÕs Sixth Framework Programme (LSHM-CT-2007-037286).The authors also wish to acknowledge NASA for funding in support of NAA through the Vermont Space Grant Fellowship; Mike Schmidt, creator of the GP package Eureqa used in this study; and Ilknur Icke, Trevor Andrews and Richard Watts for crucial conceptual and technical support.

%% The Appendices part is started with the command \appendix;
%% appendix sections are then done as normal sections
\appendix

\section{\label{sec:ROI}Table of ROI}
% !TEX root =  InteractionHierarchy.tex
%\renewcommand{\arraystretch}{2}
\setlength{\fboxsep}{2pt}%
\setlength{\fboxrule}{0pt}

In Table \ref{tab:ROI} we list all 52 ROI investigated in this study by number, give their anatomical names, indicate the ICN from within which they were defined, and provide visual representations of their locations within the brain.  Due to its length, the table appears after the References.

\begin{table*}[ht]
\centering
\caption{\label{tab:ROI}Table of ROI} \vspace{.1in}
\begin{tabular}{|M{.04\textwidth}|M{.04\textwidth}|M{.38\textwidth}|M{.4\textwidth}|}\hline
  ROI  & ICN & ROI Description & Visualization  \\ \hline
  1 & 1 & left anterior hippocampus & \multirow{2}{*}{\fbox{\includegraphics[trim=0cm 4cm 0cm 4cm, clip=true, width=.3\textwidth]{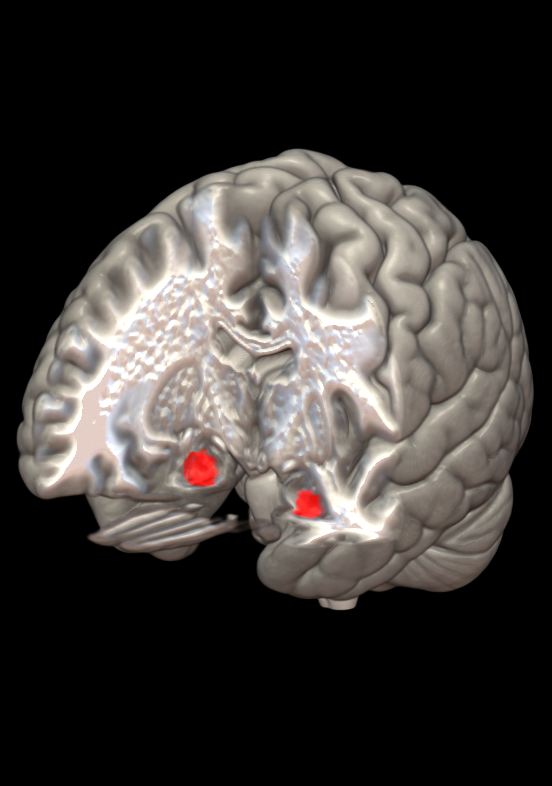}}}  \\[70pt] \cline{1-3}
  2 & 1 & right anterior hippocampus &   \\[70pt] \hline
  3 & 2 & subgenual anterior cingulate cortex, anterior caudate & \multirow{2}{*}{\fbox{\includegraphics[trim=0cm 4cm 0cm 4cm, clip=true, width=.3\textwidth]{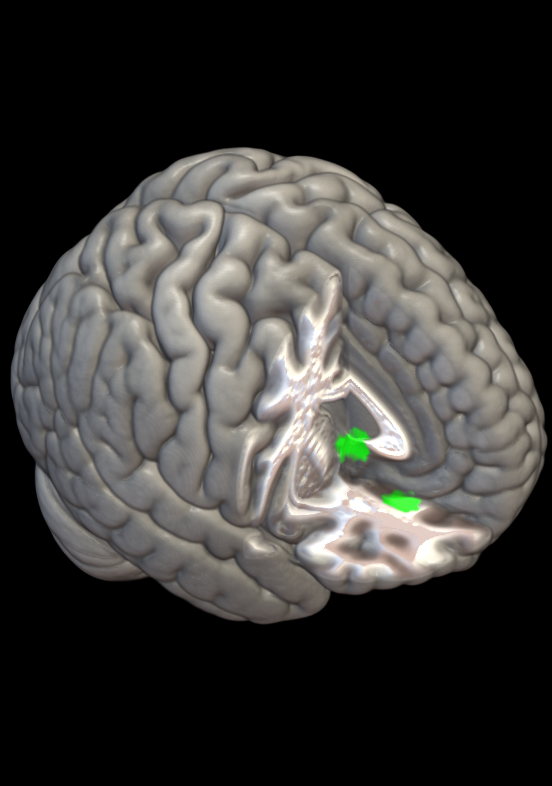}}}  \\[70pt] \cline{1-3}
  4 & 2 & ventromedial prefrontal cortex, medial frontal gyrus &  \\[70pt] \hline
  5 & 3 & right globus pallidus & \multirow{2}{*}{\fbox{\includegraphics[trim=0cm 5cm 0cm 4cm, clip=true, width=.3\textwidth]{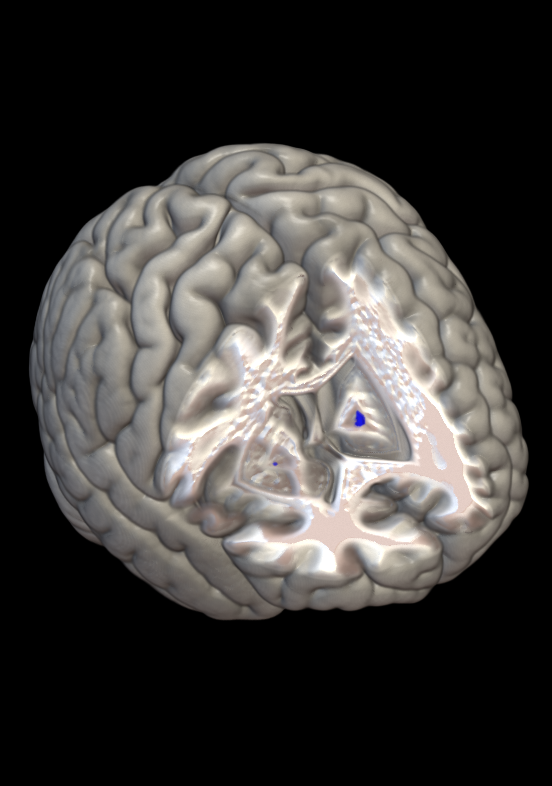}}}  \\[66pt] \cline{1-3}
  6 & 3 & left globus pallidus &   \\[66pt] \hline
  7 & 4 & right anterior insula & \multirow{3}{*}{\fbox{\includegraphics[trim=0cm 4.5cm 0cm 4cm, clip=true, width=.3\textwidth]{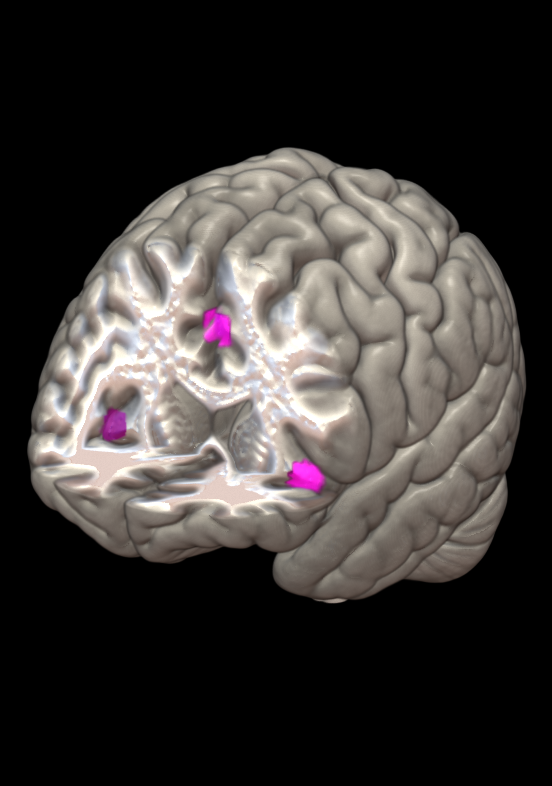}}}  \\[41pt] \cline{1-3}
  8 & 4 & left anterior insula &   \\[41pt] \cline{1-3}
  9 & 4 & middle cingulate cortex, dorsomedial prefrontal cortex &   \\[41pt] \hline
\end{tabular}
\end{table*}

\begin{table*}[ht]
\centering
\begin{tabular}{|M{.04\textwidth}|M{.04\textwidth}|M{.38\textwidth}|M{.4\textwidth}|}\hline
  ROI  & ICN & ROI Description & Visualization  \\ \hline
  10 & 5 & inferior cerebellum & \multirow{3}{*}{\fbox{\includegraphics[trim=0cm 4cm 0cm 4.5cm, clip=true, width=.3\textwidth]{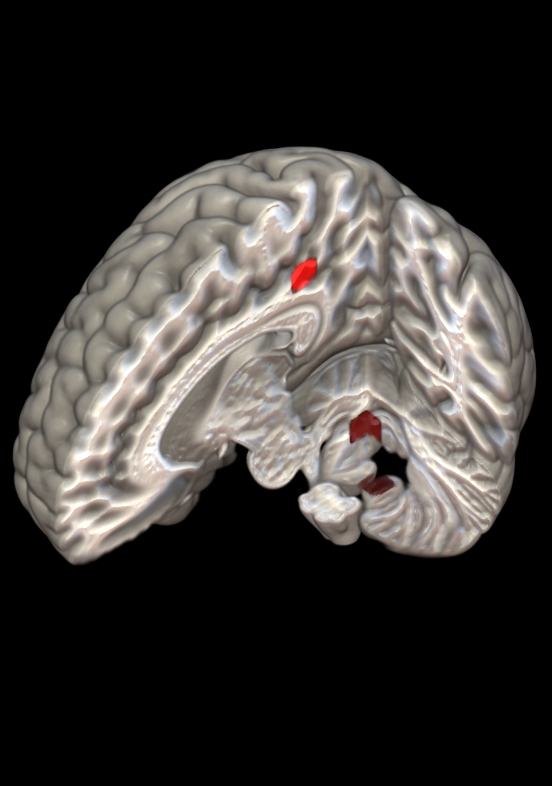}}}  \\[41pt] \cline{1-3}
  11 & 5 & posterior cingulate cortex &   \\[41pt] \cline{1-3}
  12 & 5 & inferior vermis &   \\[41pt] \hline
  13 & 5 & inferior vermis & \multirow{3}{*}{\fbox{\includegraphics[trim=0cm 4cm 0cm 4.5cm, clip=true, width=.3\textwidth]{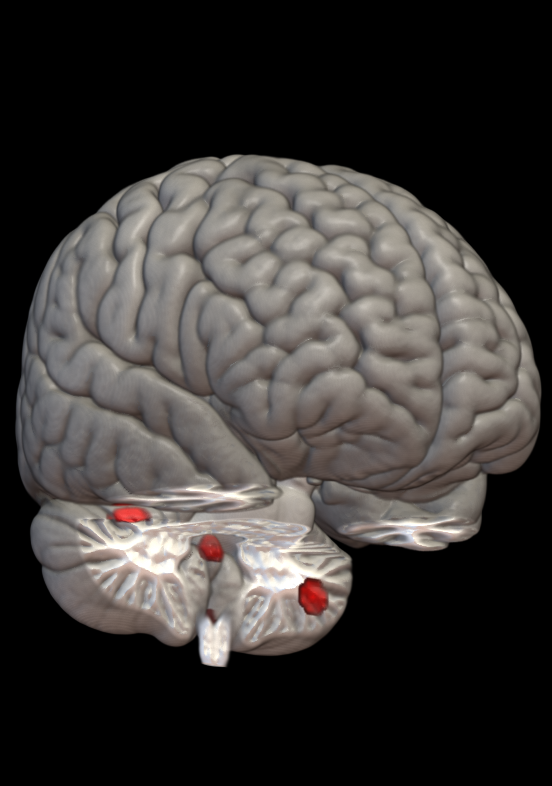}}}  \\[41pt] \cline{1-3}
  14 & 5 & anterolateral cerebellum &   \\[41pt] \cline{1-3}
  15 & 5 & anterolateral cerebellum &   \\[41pt] \hline
  16 & 5 & posterior cerebellum & \multirow{3}{*}{\fbox{\includegraphics[trim=0cm 4cm 0cm 4.5cm, clip=true, width=.3\textwidth]{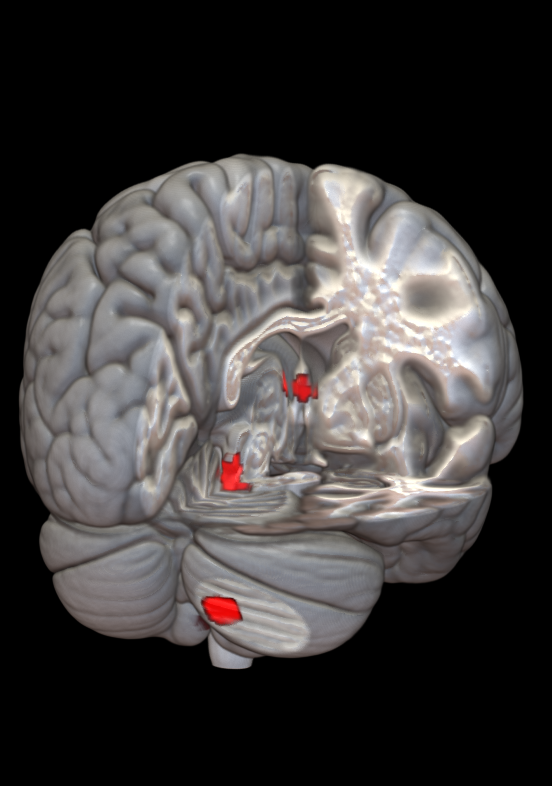}}}  \\[41pt] \cline{1-3}
  17 & 5 & inferior colliculus, anterior vermis &   \\[41pt] \cline{1-3}
  18 & 5 & fornix (body) &   \\[41pt] \hline
  19 & 6 & posterior dorsomedial prefrontal cortex & \multirow{3}{*}{\fbox{\includegraphics[trim=0cm 4cm 0cm 4.5cm, clip=true, width=.3\textwidth]{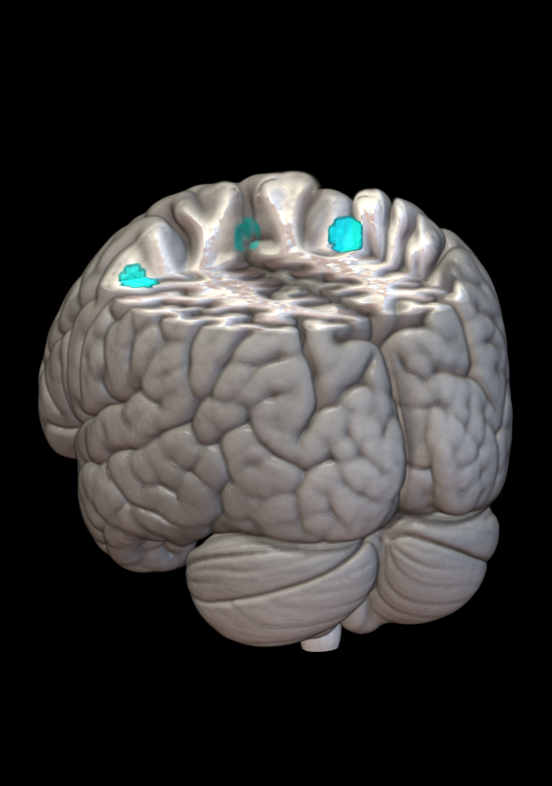}}}  \\[41pt] \cline{1-3}
  20 & 6 & left superior precentral gyrus &   \\[41pt] \cline{1-3}
  21 & 6 & right posterior superior parietal cortex &   \\[41pt] \hline
\end{tabular}
\end{table*}

%\begin{table*}[ht]
%\centering
%\begin{tabular}{|M{.04\textwidth}|M{.04\textwidth}|M{.38\textwidth}|M{.4\textwidth}|}\hline
%  ROI  & ICN & ROI Description & Visualization  \\ \hline
%\end{tabular}
%\end{table*}

\begin{table*}[ht]
\centering
\begin{tabular}{|M{.04\textwidth}|M{.04\textwidth}|M{.38\textwidth}|M{.4\textwidth}|}\hline
  ROI  & ICN & ROI Description & Visualization  \\ \hline
  22 & 7 & left superior parietal cortex & \multirow{6}{*}{\fbox{\includegraphics[trim=0cm 3cm 0cm 4.5cm, clip=true, width=.3\textwidth]{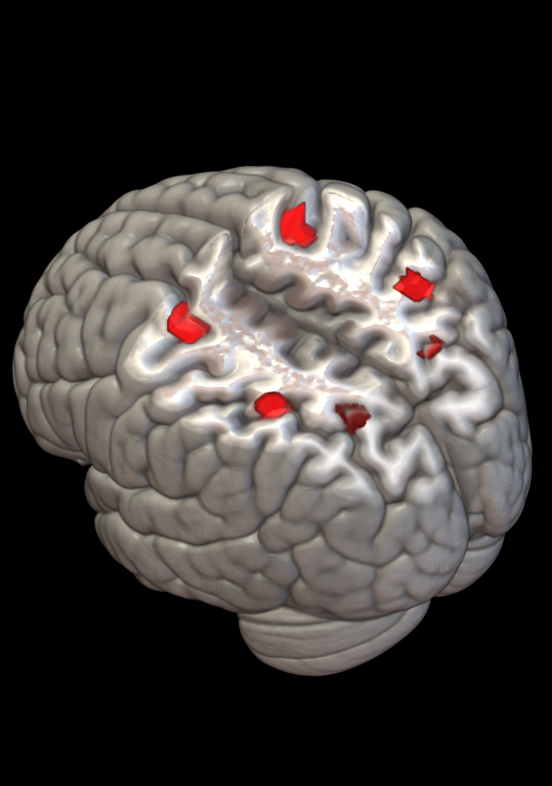}}}  \\[16pt] \cline{1-3}
  23 & 7 & right precuneus &   \\[16pt] \cline{1-3}
  24 & 7 & left superior parietal cortex &   \\[16pt] \cline{1-3}
  25 & 7 & right superior parietal cortex &   \\[16pt] \cline{1-3}
  26 & 7 & left posterior dorsolateral prefrontal cortex &   \\[16pt] \cline{1-3}
  27 & 7 & right posterior dorsolateral prefrontal cortex  &   \\[16pt] \hline
  28 & 8 & left postcentral gyrus & \multirow{4}{*}{\fbox{\includegraphics[trim=0cm 4cm 0cm 4.5cm, clip=true, width=.3\textwidth]{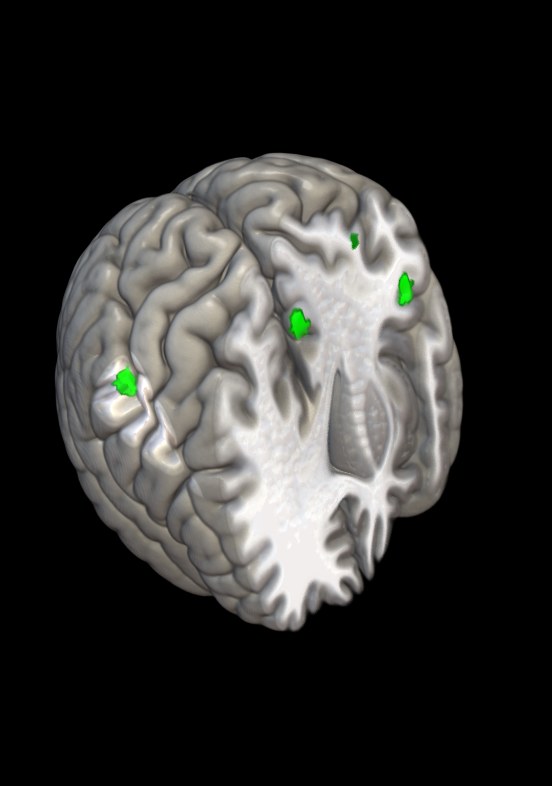}}}  \\[28pt] \cline{1-3}
  29 & 8 & paracentral lobule &   \\[28pt] \cline{1-3}
  30 & 8 & anterior inferior parietal cortex &   \\[28pt] \cline{1-3}
  31 & 8 & right postcentral gyrus &   \\[28pt] \hline
  32 & 10 & right lateral occipital cortex & \multirow{4}{*}{\fbox{\includegraphics[trim=0cm 3.5cm 0cm 5cm, clip=true, width=.3\textwidth]{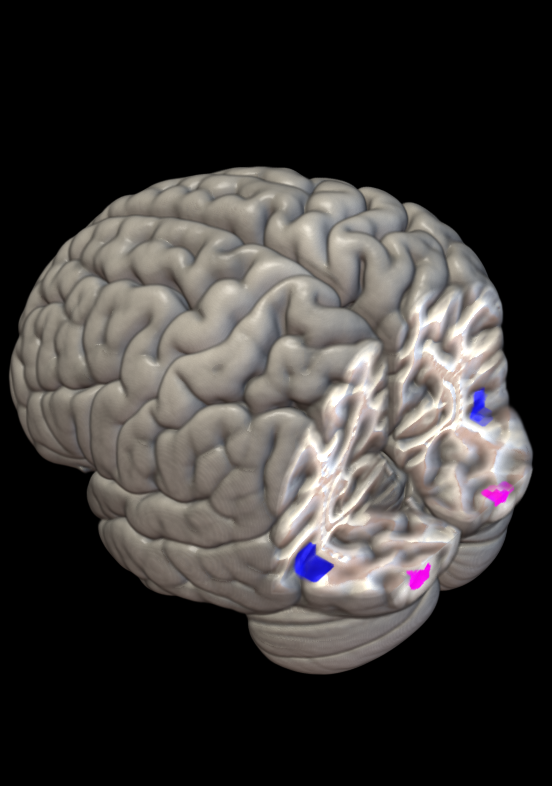}}}  \\[28pt] \cline{1-3}
  33 & 10 & left lateral occipital cortex &   \\[28pt] \cline{1-3}
  34 & 11 & left occipital pole &   \\[28pt] \cline{1-3}
  35 & 11 & right occipital pole &   \\[28pt] \hline
  36 & 12 & lingual gyrus & \multirow{3}{*}{\fbox{\includegraphics[trim=0cm 4cm 0cm 4.5cm, clip=true, width=.3\textwidth]{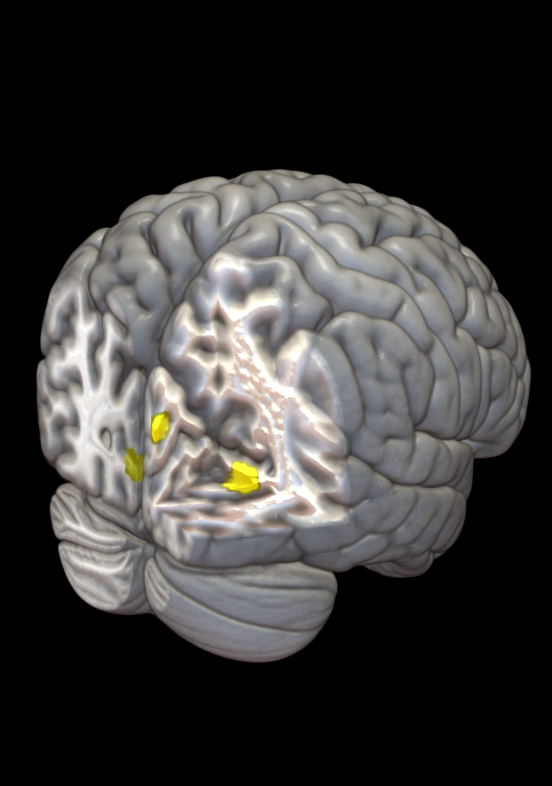}}}  \\[41pt] \cline{1-3}
  37 & 12 & right cuneus &   \\[41pt] \cline{1-3}
  38 & 12 & right fusiform gyrus &   \\[41pt] \hline
\end{tabular}
\end{table*}

\begin{table*}[ht]
\centering
\begin{tabular}{|M{.04\textwidth}|M{.04\textwidth}|M{.38\textwidth}|M{.4\textwidth}|}\hline
  ROI  & ICN & ROI Description & Visualization  \\ \hline
  39 & 13 & posterior cingulate cortex & \multirow{4}{*}{\fbox{\includegraphics[trim=0cm 4cm 0cm 4.5cm, clip=true, width=.3\textwidth]{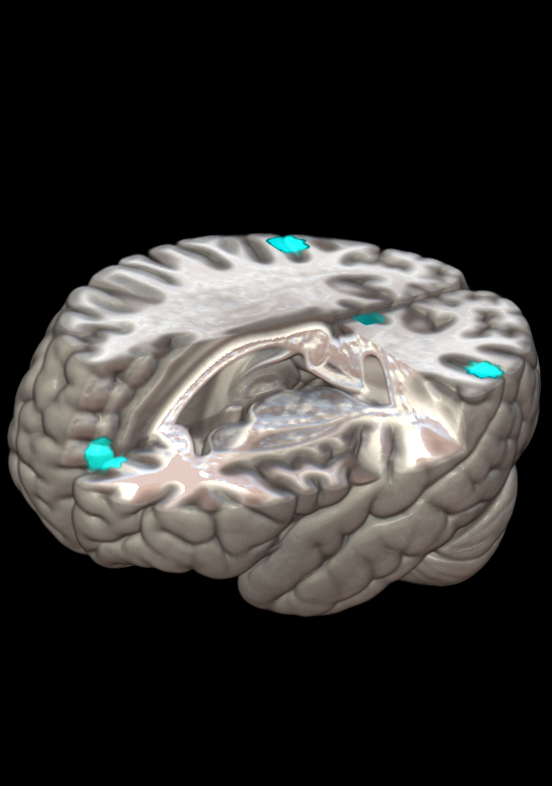}}}  \\[28pt] \cline{1-3}
  40 & 13 & left angular gyrus &   \\[28pt] \cline{1-3}
  41 & 13 &  right angular gyrus &   \\[28pt] \cline{1-3}
  42 & 13 & anterior dorsomedial prefrontal cortex &   \\[28pt] \hline
  43 & 14 & right superior cerebellum & \multirow{3}{*}{\fbox{\includegraphics[trim=0cm 4cm 0cm 4.5cm, clip=true, width=.3\textwidth]{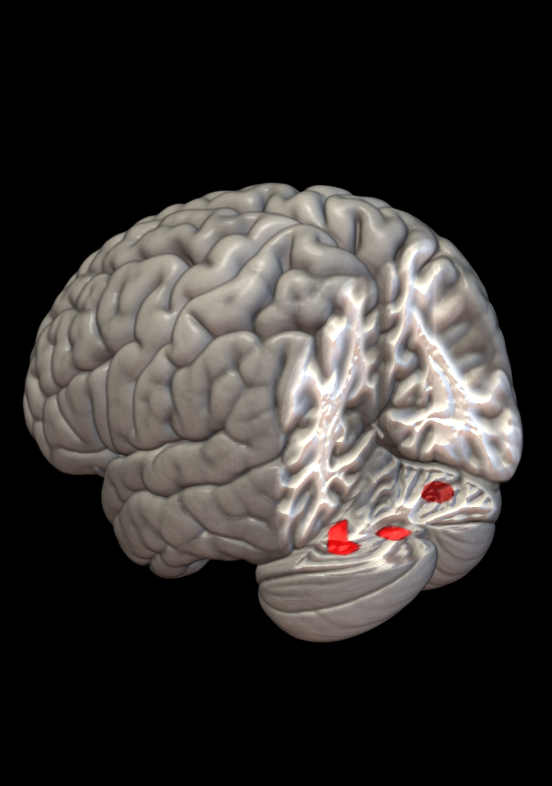}}}  \\[41pt] \cline{1-3}
  44 & 14 & vermis &   \\[41pt] \cline{1-3}
  45 & 14 & left superior cerebellum &   \\[41pt] \hline
  46 & 15 & right middle frontal gyrus & \multirow{2}{*}{\fbox{\includegraphics[trim=0cm 5cm 0cm 4cm, clip=true, width=.3\textwidth]{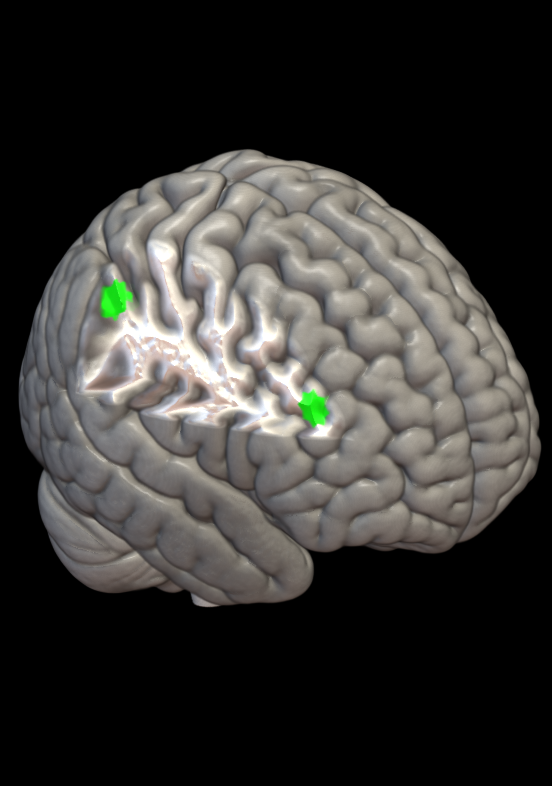}}}  \\[66pt] \cline{1-3}
  47 & 15 & right supramarginal gyrus &   \\[66pt] \hline
  48 & 16 & left superior temporal gyrus & \multirow{2}{*}{\fbox{\includegraphics[trim=0cm 5cm 0cm 4cm, clip=true, width=.3\textwidth]{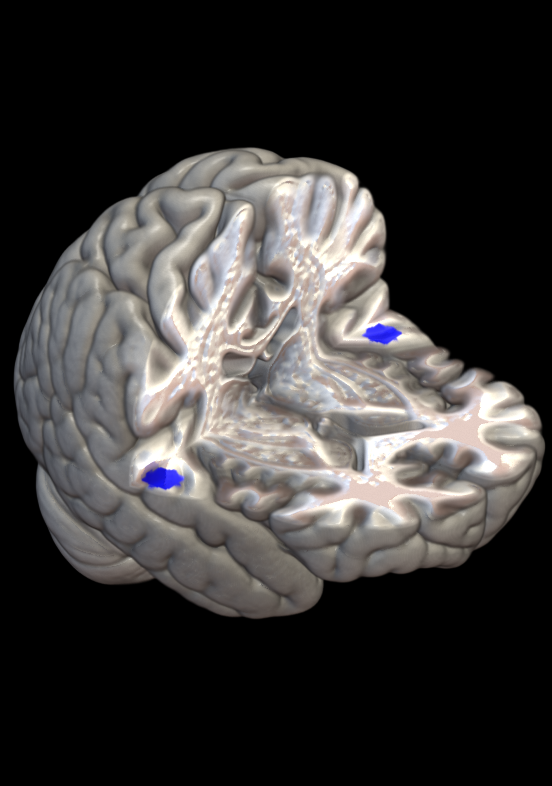}}}  \\[66pt] \cline{1-3}
  49 & 16 & right superior temporal gyrus &   \\[66pt] \hline %\cline{1-3}
\end{tabular}
\end{table*}

\begin{table*}[ht]
\centering
\begin{tabular}{|M{.04\textwidth}|M{.04\textwidth}|M{.38\textwidth}|M{.4\textwidth}|}\hline
  ROI  & ICN & ROI Description & Visualization  \\ \hline
  50 & 17 & right inferior pre and post central gyrus & \multirow{3}{*}{\fbox{\includegraphics[trim=0cm 4cm 0cm 4.5cm, clip=true, width=.3\textwidth]{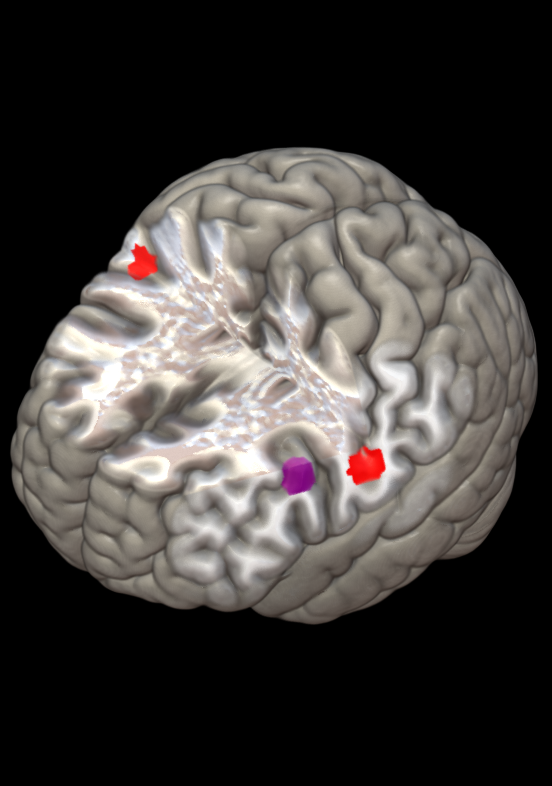}}}  \\[41pt] \cline{1-3}
  51 & 17 & left inferior pre and post central gyrus &   \\[41pt] \cline{1-3}
  52 & 18 & left inferior frontal gyrus &   \\[41pt] \hline
\end{tabular}
\end{table*}

\section{\label{sec:indvar}Subject-level variation}
\begin{figure*}[htp]
\centering
\subfigure{\includegraphics[width=\textwidth]{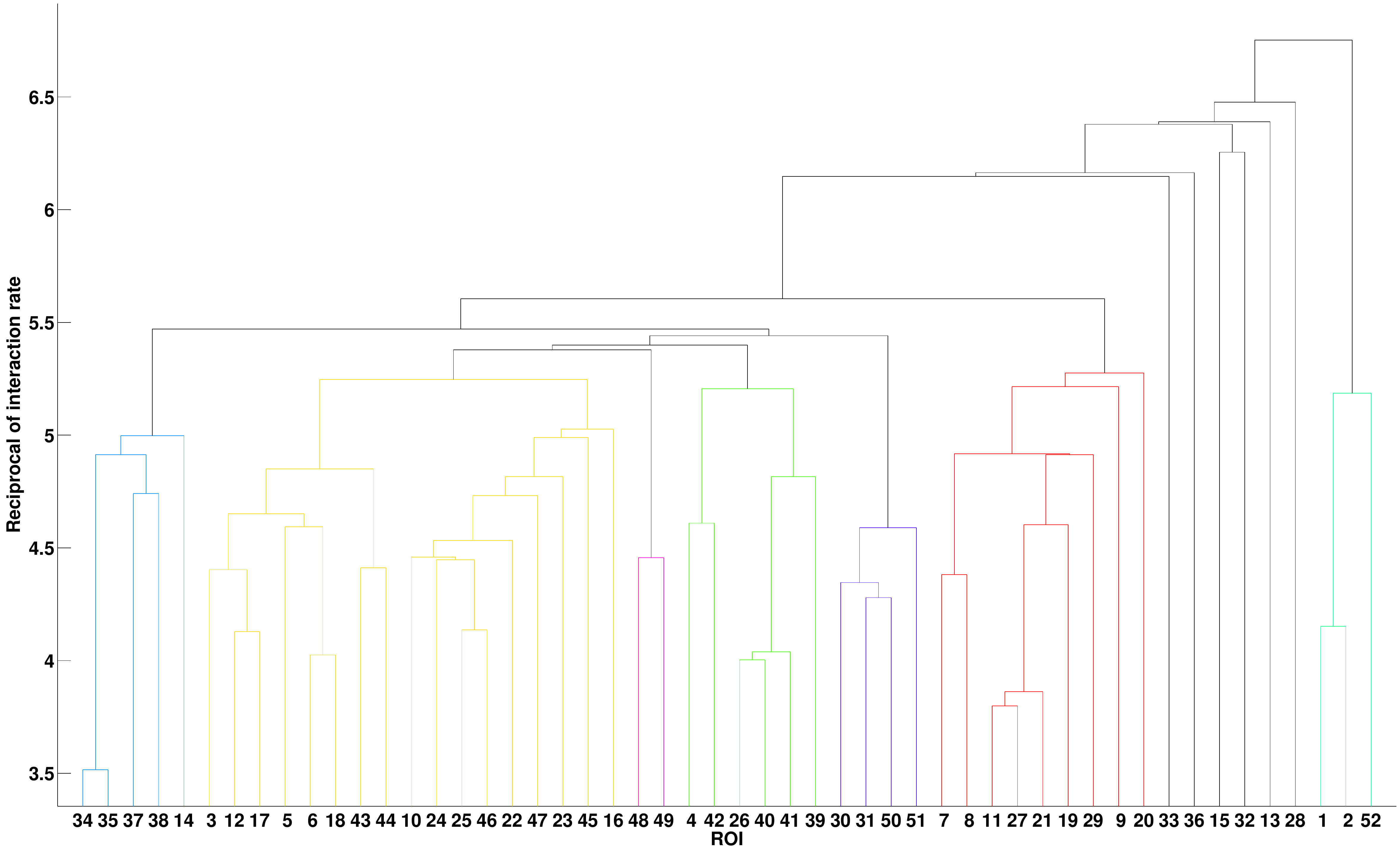}}\\
\subfigure{\includegraphics[width=\textwidth]{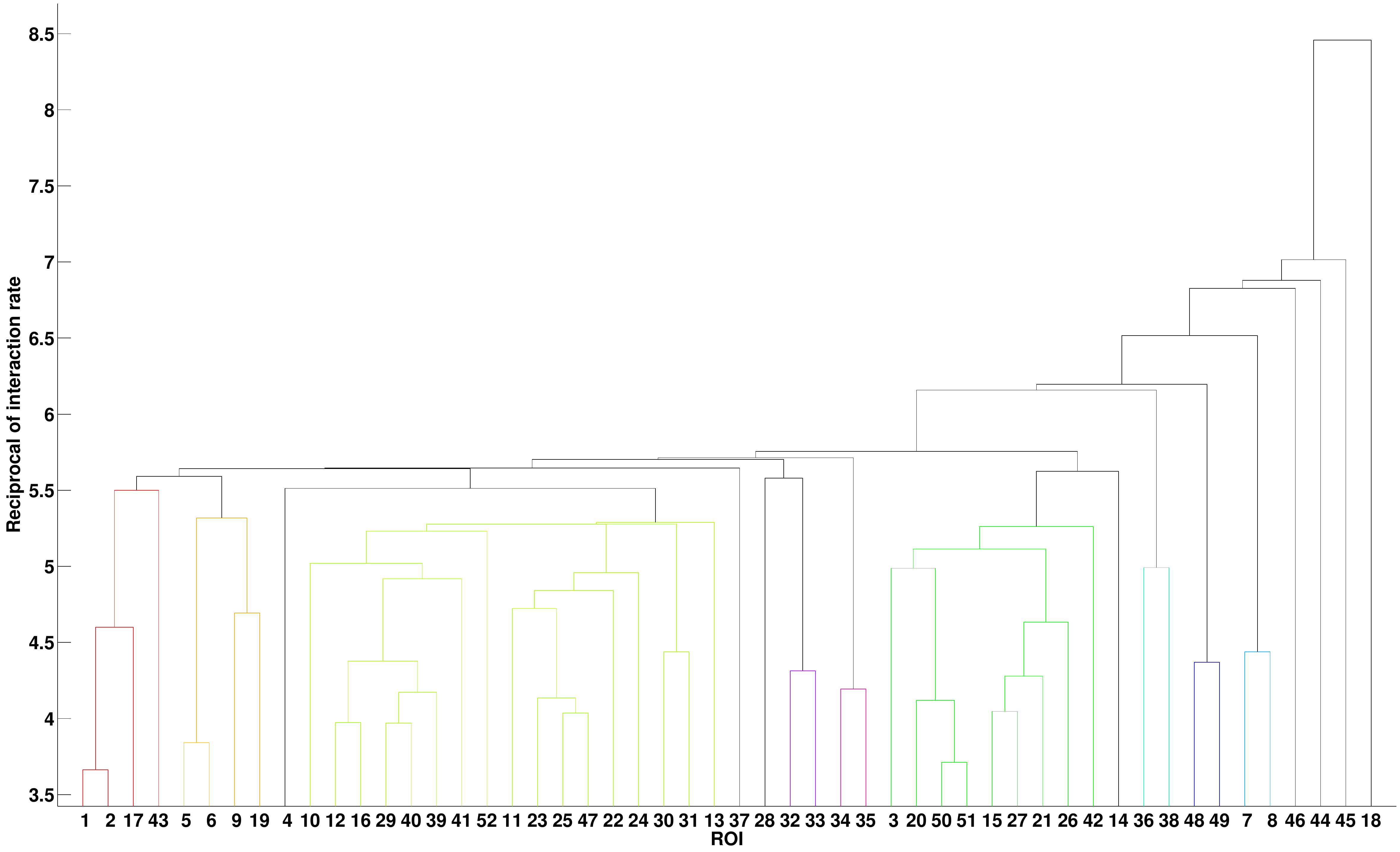}}
\caption[Example hierarchies for two different individual subjects]{\label{fig:dendroind}Example hierarchies for two different individual subjects. Note the large degree of variation between the two.}
\end{figure*}

Figure \ref{fig:dendroind} contains individual subject interaction hierarchies for two different (randomly selected) subjects, generated by 100 random restarts of the GP algorithm and subsequent normalized frequency analysis.  Though the two hierarchies are quite different from one another, they do show some network organization similar to that illustrated in the population level hierarchy (top of Figure \ref{fig:dendrolinnonlin}).  
\begin{itemize}
\item Portions of the visual cluster (ROI 32-38) are intact in each case. 
\item Many of the two-region networks remain together, e.g. ICN 1 (ROI 1,2), ICN 16 (ROI 48,49), and though associated with other ROI, also ICN 3 (ROI 5,6) and ICN 17 (ROI 50,51) 
\item The default mode network (ROI 39-42) is mostly intact in each subject.
\end{itemize}  

The interaction profile of the default mode network illustrates an interesting distinction between the two subjects.  For the top subject, the network is fully intact, interacting with ROI 4 (consistent with the population level hierarchy), and also interacting with ROI 26 from the motor visuospatial complex. For the bottom subject, three of the four ROI in the network remain together, but interact instead with several other ROI from the emotional interoceptive class instead of ROI 4, specifically ROI 10, 12, and 16, and a different ROI from the motor visuospatial complex as well (ROI 29 instead of 26).  By themselves, these dendrogram comparisons offer no conclusive evidence regarding connections between cognitive processes. However, an experiment could be designed to test if any inferences can be made from such distinctions.  For example, the administration of post-scan surveys might grant some interpretability to the specifics of these single-subject interaction hierarchies.   

\section{\label{sec:alclin}Linear HCA of alcohol consumption}
\begin{figure*}[htp]
\centering
\subfigure{\includegraphics[width=\textwidth]{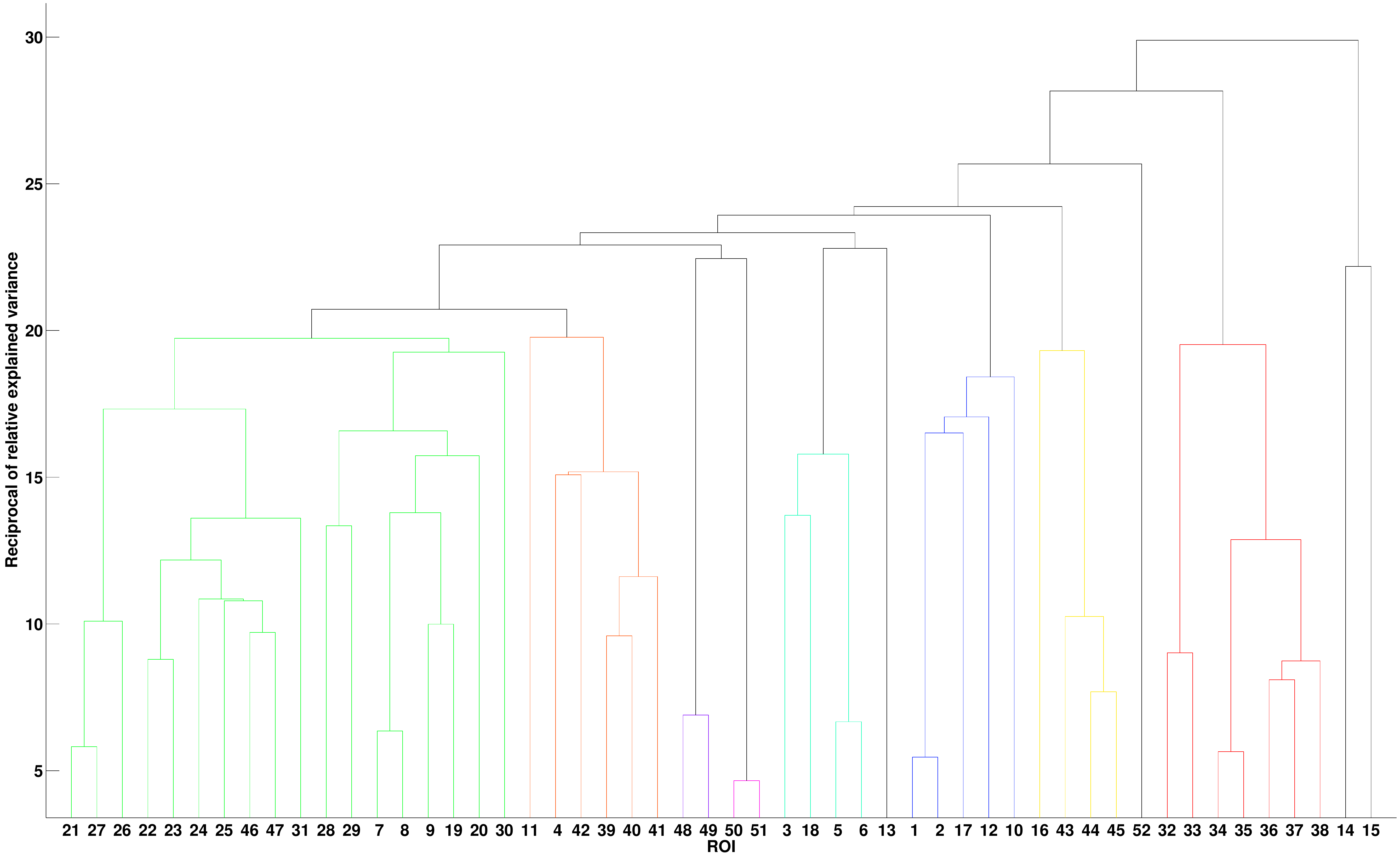}}\\
\subfigure{\includegraphics[width=\textwidth]{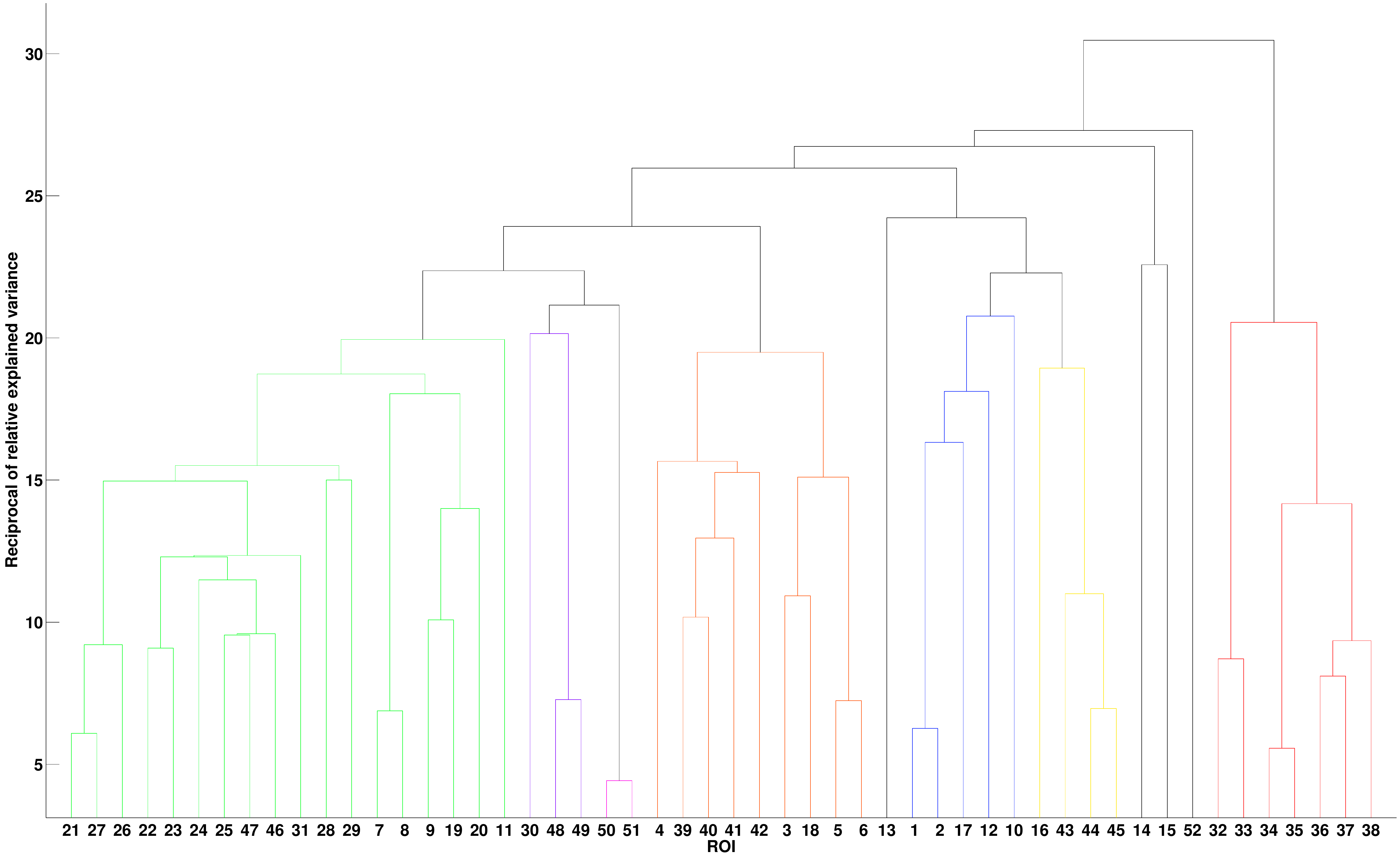}}
\caption[Linear hierarchies for groups with high and low alcohol consumption rates]{\label{fig:dendroalclin}Linear hierarchies for groups with high (top) and low (bottom) alcohol consumption rates, defined by two or more lifetime drinks and one or fewer lifetime drinks, respectively.}
\end{figure*}

Here we demonstrate that the shuffling of the interaction hierarchy in drinking (D) versus non-drinking (ND) adolescents discovered by NFM is not uncovered by linear correlation analysis.  To perform group-level correlation analysis, the normalized relative $R^2$ matrices for each subject (described in Section \ref{sec:linvnonlin}) are averaged over the 100 subjects in each group.  Recall that these matrices are generated for each subject by computing the correlation matrix for the 52 ROI time series, squaring the elements, and normalizing each row (after setting the diagonal to zero). The reciprocal of relative explained variance can be considered a distance between ROI (higher relative $R^2$ means closer), and the resulting D and ND hierarchies generated by HCA are shown in the top and bottom, respectively, of Figure \ref{fig:dendroalclin}.

Comparison with Figure \ref{fig:dendroalc} suggests that this linear analysis \textsl{partially} uncovers a distinguishing difference in interaction between drinking and non-drinking adolescents.  Specifically, among non-drinkers, a higher intra-network interaction within ICN 2, comprised of the subgenual ACC and the vmPFC (ROI 3 and 4, respectively) is detected here.  The result is an indirect coupling, within non-drinkers, of the default mode network (ROI 39-42) and the complex comprised of ROI 3,18,5,6, through the vmPFC.  

The results of NFM provide further insight in two important ways. First, the elevated interaction within ICN 2 among non-drinkers is detected at twice the strength.  Second, the main interaction responsible for grouping the complex of ROI 3,18,5,6, specifically the interaction between the left globus pallidus (ROI 6) and fornix body (ROI 18), is lower among non-drinkers. This second effect is entirely missed by correlation analysis, suggesting that it is nonlinear in nature.  The result of capturing these effects \textsl{together}, as shown in Figure \ref{fig:dendroalc}, is a breakup of the complex in non-drinkers, for whom ROI 3,18 are separated from the bilateral globus pallidus of ICN 3 (ROI 5-6).  This breakup is suggestive, as each of these ROI is associated with emotion, reward, and interoceptive processes such as thirst, and experiments reporting activity in ICN 5, including the fornix body (ROI 18), predominantly involved interoceptive stimulation, as reported in \cite{laird2011}.

\section{\label{sec:improvements}Improvements and modifications}
The GP implementation we used for this study is the commercially available package Eureqa from Nutonian, as described in \cite{schmidt2009}.  Though much of its behavior can be controlled through the interface or command line, it is proprietary code and thus somewhat of a black box.  There are many reasons why a dedicated, open source implementation of GP would be more desirable. 

A major challenge for this method of analysis is the computational expense of running a large number of GP searches.  Generating the IR map for a single subject requires a large number of random restarts for each ROI.  For example, running 100 restarts for each of the 52 ROI in this study, allowing 1 core-hour for each search, requires over 10 hours with access to 500 dedicated processors.  The procedure as described here is likely computationally prohibitive for running analyses on large numbers of subjects, or for larger collections of ROI.  Intelligent stopping criteria, and many other approaches to the mitigation of computational expense, have been reported at length in the GP literature, an example of which is the use of graphics processors reported in \cite{harding2007}.  It may also be possible to determine an ideal (and smaller) number of restarts that balances computation time with the statistical power of the resulting IR map.  

It should also be noted that for collections of ROI much larger than that considered here, in addition to the computational expense resulting from more required searches, each search will take much longer to produce meaningful models due to the larger number of possible explanatory variables.  A hybrid method of symbolic regression employing a machine learning algorithm called FFX (Fast Function Extraction) described in \cite{mcconaghy2011} as a first pass, and then GP, has great potential for the treatment of higher dimensional data, e.g., large numbers of ROI.  A prototype of this method was reported in \cite{icke2014}. FFX is a deterministic algorithm that builds up models with nonlinear terms (e.g., products of ROI signal) in a prescribed fashion and evaluates explanatory power at each stage.  By ruling out ROI that are likely not explanatory at each stage, the algorithm reduces the dimensionality of the search.  In other words, at the cost of reduced breadth in the search space, the algorithm provides huge reductions in computation time in addition to reducing the number of variables that will eventually be injected into the GP algorithm.  Implemented effectively, this hybrid algorithm could eliminate the necessity of ROI selection completely by allowing direct regression over voxel signals.  

An ever-present concern in the analysis of fMRI is the level of noise in the data.  Particularly in the case of regressing over voxel signals, low signal-to-noise ratio is a major challenge, and indeed GP efficacy is diminished in such circumstances.  However, there has been some work on modifying the GP algorithm to better manage noisy data, an example of which is the inclusion of noise generators called \textsl{stochastic elements} with user-defined distributions (e.g., Gaussian or uniform) as potential explanatory ``variables''.  These generators can themselves end up inside complex functions within the models, providing those models the capability of reproducing realistic noise distributions more likely to be at play than the typical Gaussian.  There is no guarantee that this modification will prove beneficial in the case of fMRI, but it has been shown, in \cite{schmidt2007}, to effectively identify exact underlying analytical models in the presence of nonlinear, non-Gaussian and nonuniform noise.

\bibliographystyle{elsarticle-harv}
\bibliography{bibfile.bib}

%% Authors are advised to submit their bibtex database files. They are
%% requested to list a bibtex style file in the manuscript if they do
%% not want to use elsarticle-harv.bst.

%% References without bibTeX database:

% \begin{thebibliography}{00}

%% \bibitem must have one of the following forms:
%%   \bibitem[Jones et al.(1990)]{key}...
%%   \bibitem[Jones et al.(1990)Jones, Baker, and Williams]{key}...
%%   \bibitem[Jones et al., 1990]{key}...
%%   \bibitem[\protect\citeauthoryear{Jones, Baker, and Williams}{Jones
%%       et al.}{1990}]{key}...
%%   \bibitem[\protect\citeauthoryear{Jones et al.}{1990}]{key}...
%%   \bibitem[\protect\astroncite{Jones et al.}{1990}]{key}...
%%   \bibitem[\protect\citename{Jones et al., }1990]{key}...
%%   \harvarditem[Jones et al.]{Jones, Baker, and Williams}{1990}{key}...
%%

% \bibitem[ ()]{}

% \end{thebibliography}

\end{document}